\documentclass[a4paper,dvips,12pt]{article}
\usepackage{amsmath}

\def\slash{\not{\! \! \rm}}
\def\id{\ 1 \! \! \! \! 1}
\def\Tr{\rm Tr}

\title{
\vspace*{-0.8cm}
\begin{flushright}
\normalsize{CERN--PH--TH/2004-099\\
\texttt{hep-th/0406010}}\\
\end{flushright}
\vspace{1cm}
\Large\textbf{Non-linear supersymmetry and intersecting D-branes}
\author{\large
{\bf I.~Antoniadis$^1$\footnote{On leave of absence from CPHT,
Ecole Polytechnique, UMR du CNRS 7644.},
M.~Tuckmantel$^{1,2}$}\\ \\
\emph{$^1$Department of Physics, CERN - Theory Division}\\
\emph{CH--1211 Geneva 23, Switzerland}\\
\emph{$^2$Institut f\"ur Theoretische Physik, ETH H\"onggerberg}\\
\emph{CH--8093\, Z\"urich, Switzerland}}}
\date{}
\begin{document}
\maketitle
\thispagestyle{empty}
\vspace*{.5cm}

\begin{abstract}
We study the non-linear realization of supersymmetry.
We classify all lower dimensional operators, describing effective
interactions of the Goldstino with Standard Model fields. Besides a
universal coupling to the energy momentum tensor of dimension eight,
there are additional model dependent operators whose strength is not
determined by non-linear supersymmetry, within the effective field
theory. Their dimensionality can be lower than eight, starting with
dimension six, leading in general to dominant effects at low energies.
We compute their coefficients in string models with D-branes at angles.
We find that the Goldstino decay constant is given by the total brane
tension, while the various dimensionless couplings are independent from
the values of the intersection angles.
\end{abstract}
\date
\newpage

\section{Introduction} \label{introduction}

In this work, we study the non-linear realization of supersymmetry,
present in a class of D-brane models with non supersymmetric spectra.
In fact generic D-brane configurations in type II closed superstring
theories~\cite{Angelantonj:2002ct}, either combined with orientifolds~\cite{bsb} or at
angles~\cite{Berkooz:1996km, bi, mf}, break all bulk supersymmetries which
are however still realized on the D-branes world-volume in a non-linear
way~\cite{Volkov:ix,Bagger:1996wp}. A consequence of non-linear
realization is the existence of a (tree-level) massless Goldstino(s)
which is a brane field~\cite{Bagger:1996wp,bnl,Antoniadis:2001pt}. In the
compact case, it is expected to acquire a small mass by radiative
corrections, suppressed by the compactification
volume~\cite{Antoniadis:2001pt,Antoniadis:2004qn}. Our analysis is of
particular interest for models where the string scale is in the TeV
region and supersymmetric bulk~\cite{Antoniadis:1998ig}, or even in
models with light Goldstino and all superparticles heavier than the
electroweak
scale~\cite{Clark:1996aw,Brignole:1997pe,Luty:1998np,Clark:1998aa}.

Our aim is to determine the effective action describing the lower
dimensional interactions of the Goldstino with all kinds of fields: gauge
bosons, scalars and (chiral) fermions. It is known
that non-linear supersymmetry implies a universal coupling between the
Goldstino and matter stress-energy tensors of dimension eight, whose
strength is fixed by the Goldstino decay constant, in analogy to
low-energy theorems for spontaneously broken global symmetries.
However, it was noticed that besides this coupling there may exist
other supersymmetric interactions whose strength is left undetermined
within the effective field
theory~\cite{Brignole:1997pe,Luty:1998np,Clark:1998aa}. For instance, a
general analysis of Goldstino to fermions interactions, which are
described to lowest order by dimension eight four-fermion operators,
revealed the existence of a free parameter, associated to the coefficient
of a second operator allowed by non-linear supersymmetry, besides the one
corresponding to  the product of the two stress-tensors. This parameter
can be computed in principle in string theory by considering the low
energy expansion of appropriate four-fermion
amplitudes~\cite{Antoniadis:2000jv,Antoniadis:2001pt}.

In this work, we extend the general analysis of Goldstino interactions
to gauge and scalar fields and compute the leading coefficients in
string theory. At the four-point level, we find for instance
additional operators of dimension eight, involving two scalars and two
Goldstinos, similar to the four-fermion operator described above.
However, now there are more interactions of lower dimensionality. In
particular, there are two operators of dimension six that contain a
single Goldstino coupled to a matter fermion and a gauge or scalar field.
The presence of these operators complicates the extraction of the
Goldstino effective action from four-point string amplitudes because they
generate reducible contact terms that have to be subtracted. Fortunately,
such terms are absent for generic brane intersection angles, allowing to
compute the coefficients of all lower dimensional four- and three-point
vertices in the effective action. Thus, although four-point contact terms
appear to depend on intersection angles, the coefficients of irreducible
effective operators turn out to be model independent constants.

More precisely, we study the non-linear supersymmetry present locally
on the intersection of two sets of coincident D6-branes. The
intersection is point-like in the six-dimensional internal (compact)
manifold and extends in our three space non-compact coordinates. The
Goldstino is a linear combination of the two gauge singlet fermions
localized, respectively, on the two stacks. Moreover, its decay
constant is given by the effective total brane tension on the intersection. We
then determine its leading interactions with the massless gauge and matter
fields living on the two sets, as well as with the chiral fermions and
scalars localized at the intersection.

Our paper is organized as follows. The next three sections (2-4)
are devoted to the effective field theory. In Section 2, we recall the
main properties of non-linear supesrymmetry, such as field
transformations  and invariant actions. In Section 3, we describe the
superfield formalism, while in Section 4, we derive all supersymmetric
Goldstino couplings to gauge fields, scalars and fermions, up to
dimension eight. In the following three sections (5-7), we determine all
three- and four-point couplings among brane fields. In Section 5, we
compute all four-point functions on a disc world-sheet, involving two
Goldstinos and two  matter fields from a single set of coincident
D-branes. In Section 6, we generalize this computation for matter fields
living on brane intersections.  For particular values of intersection
angles, there are additional massless gauge or scalar fields that have
3-point  interactions with Goldstinos leading to reducible contributions
in the 4-point functions. By studying the various degeneration limits,
we extract the 3-point interactions. In Section~7, we combine the
previous results and determine the strength of all effective operators
describing the lowest dimensional Goldstino interactions, up to
four-point level. Finally, Section 8 contains our conclusions. For
convenience of the reader, there are also two appendices with a summary
of our results. Appendix A contains the Goldstino couplings from the
general analysis of non-linear supersymmetry, while Appendix B contains
the low-energy effective action derived from the string computations.

\section{Non-linear supersymmetry}
\setcounter{equation}{0}

In the standard realization of non-linear supersymmetry, the
Goldstino field $\lambda$ transforms according to

\begin{eqnarray}
\delta \lambda_{\alpha} = \frac{\xi_{\alpha}}{\kappa} - i \kappa (
\lambda \sigma^{\mu} \bar{\xi} - \xi \sigma{^\mu} \bar{\lambda} )
\partial_{\mu} \lambda_{\alpha} \nonumber \\
\delta \bar{\lambda}^{\dot{\alpha}} =
\frac{\bar{\xi}^{\dot{\alpha}}}{\kappa} - i \kappa ( \lambda
\sigma^{\mu} \bar{\xi} - \xi \sigma{^\mu} \bar{\lambda} )
\partial_{\mu} \bar{\lambda}^{\dot{\alpha}}
\label{goldstino_transformation}
\end{eqnarray}
Here $\xi_{\alpha}$,\ $\bar{\xi}^{\dot{\alpha}}$ are the
(Grassmann) parameters of the transformation and $\kappa$ is some
constant with units of $length^{2}$ which parametrizes  the
supersymmetry breaking scale. It is the Goldstino decay constant and
plays a role similar to $f_{\pi}$
in soft-pion dynamics.

We wish to construct a Lagrangian for an effective low-energy
description of the Goldstino and its interactions with Standard
Model fields. We first consider the part of the effective action
which contains only self-couplings of the Goldstino. This must
contain the standard kinetic term for a Weyl spinor with some
additional terms necessary in order to make the action invariant
under the standard non-linear realization~(\ref{goldstino_transformation}).
An action which satisfies these
criteria has been constructed by Akulov and Volkov~\cite{Volkov:ix}.
Indeed, we define the
quantity
\begin{equation}
A_{\mu}^{\ \nu} = \delta_{\mu}^{\ \nu} + i \kappa^{2} \lambda
\overset{\leftrightarrow}{\partial_{\mu}}\sigma^{\nu}\bar{\lambda}
\label{Amatrix}
\end{equation}
from which we can construct the Akulov-Volkov action
\begin{equation}
S_{AV} = \int d^{4}x\ \mathcal{L}_{AV} = -\frac{1}{2 \kappa^2}
\int d^{4}x\ detA\, .
\end{equation}

Using~(\ref{goldstino_transformation}) it can be shown that
$\mathcal{L}_{AV}$ transforms as a total divergence:
\begin{equation}
\delta(detA) = -i \kappa \partial_{\mu}[( \lambda \sigma^{\mu}
\bar{\xi} - \xi \sigma{^\mu} \bar{\lambda})detA] \equiv \kappa
\partial_{\mu}(\Lambda^{\mu}detA)\, ,
\label{detA}
\end{equation}
where we have introduced a useful short-hand notation
\begin{equation}
\Lambda^{\mu} = -i ( \lambda \sigma^{\mu} \bar{\xi} - \xi
\sigma{^\mu} \bar{\lambda})\, .
\end{equation}
This shows that the above action is supersymmetric. Expanding in
powers of $\kappa$ we obtain
\begin{equation}
\mathcal{L}_{AV} = -\frac{1}{2\kappa^2} - \frac{i}{2} \lambda
\sigma^{\mu} \overset{\leftrightarrow}{\partial_{\mu}}
\bar{\lambda} + ...
\end{equation}
where the dots denote self-couplings proportional to the second or
higher powers of $\kappa$. The first term plays the role of a
cosmological constant. As will be shown later, in string theory it is
given by the total effective tension of the D-branes. The remaining part
of the Lagrangian is just the non-linear supersymmetric extension of the
kinetic energy of a Weyl spinor.

The standard realization can be extended to matter (and gauge)
fields~\cite{Clark:1996aw,Clark:1998aa}. Let $\phi_{i}$ denote some
generic field with i an index in some representation of the Lorentz group
or of an internal symmetry group. Then we define
\begin{equation}
\delta\phi_{i} = \kappa \Lambda^{\mu} \partial_{\mu} \phi_{i}\, .
\label{field_transformation}
\end{equation}
It can be checked that this indeed provides a representation of
supersymmetry. Note that~(\ref{field_transformation}) has the same form
as the transformation of the Goldstino~(\ref{goldstino_transformation})
except for the absence of the inhomogeneous term
$\frac{\xi_{\alpha}}{\kappa}$.

As a special case, we can consider a gauge field $B_{\mu}$. By
gauge invariance, such a field can enter the Lagrangian only
through the field-strength tensor and gauge-covariant derivatives.
Both of these contain ordinary derivatives and therefore will not
transform covariantly according to the standard realization even though
$B_{\mu}$ does. This can be remedied by defining a modified field
strength-tensor:
\begin{equation}
\mathcal{F}^{a}_{\mu\nu} \equiv (A^{-1})_{\mu}^{\ \sigma}
(A^{-1})_{\nu}^{\ \rho} F^{a}_{\sigma\rho}\, ,
\label{newF}
\end{equation}
where $F^{a}_{\mu\nu}$ is the ordinary field-strength and
$(A^{-1})_{\mu}^{\ \nu}$ is the inverse of the matrix defined in~(\ref{Amatrix}). If we expand the right-hand side in powers of
$\kappa$, the first term will be $F^{a}_{\mu\nu}$, followed by
appropriate couplings to the Goldstino field. As a result, the
quantity $\mathcal{F}^{a}_{\mu\nu}$ transforms covariantly according to
the standard realization. The same procedure also works for the
covariant derivative. Starting from the ordinary gauge-covariant
derivative, we define a supersymmetry-covariant derivative according to
\begin{equation}
\mathcal{D}_{\mu}\phi_{i} \equiv (A^{-1})_{\mu}^{\ \nu}
D_{\nu}\phi_{i}\, .
\label{susy_covariant_derivative}
\end{equation}
Thus, if $\phi_{i}$ is a field transforming in the standard
realization, then so is $\mathcal{D}_{\mu}\phi_{i}$.

It is now a simple task to construct an invariant effective action. The
Standard Model Lagrangian has the form
\begin{equation}
\mathcal{L}_{SM} = \mathcal{L}_{SM}(\phi_{i}, D_{\mu}\phi_{i},
F^{a}_{\mu\nu})\, .
\end{equation}
By replacing all quantities with their SUSY-covariant counterparts
(that is $F^{a}_{\mu\nu} \rightarrow \mathcal{F}^{a}_{\mu\nu}$ and
$D_{\mu} \rightarrow \mathcal{D}_{\mu}$), the resulting Lagrangian
transforms as a field in the standard realization:
\begin{equation}
\delta \mathcal{L}_{SM}(\phi_{i}, \mathcal{D}_{\mu}\phi_{i},
\mathcal{F}^{a}_{\mu\nu}) = \kappa \Lambda^{\sigma} \partial_{\sigma}
\mathcal{L}_{SM}(\phi_{i}, \mathcal{D}_{\mu}\phi_{i},
\mathcal{F}^{a}_{\mu\nu})\, .
\end{equation}
Multiplying with $detA$ we obtain the invariant action
\begin{equation}
S_{eff} = \int d^4 x \mathcal{L}_{eff} = \int d^4x\ detA\
\mathcal{L}_{SM}(\phi_{i}, \mathcal{D}_{\mu}\phi_{i},
\mathcal{F}^{a}_{\mu\nu})\, .
\end{equation}

Indeed, using the transformation of $detA$ given in~(\ref{detA}) we see
that this action is supersymmetric. Furthermore, expanding the
effective Lagrangian $\mathcal{L}_{eff}$ in powers of $\kappa$,
the lowest ($\kappa$-independent) term is the Standard Model (SM)
Lagrangian itself. The additional terms are required to make the
action supersymmetric and describe appropriate interactions of the
SM-fields to the Goldstino. Explicitly:
\begin{equation}
\mathcal{L}_{eff} = \mathcal{L}_{SM}(\phi_{i}, D_{\mu}\phi_{i},
F^{a}_{\mu\nu}) + ( i \kappa^2 \lambda
\overset{\leftrightarrow}{\partial^{\mu}} \sigma^{\nu}
\bar{\lambda}) T_{\mu\nu} + ...
\end{equation}
where the dots denote higher powers of $\kappa$ that can be
neglected in the low-energy limit. $T_{\mu\nu}$ is the manifestly
gauge-invariant energy-momentum tensor
\begin{equation}
T^{\mu\nu} = \eta^{\mu\nu} \mathcal{L}_{SM} - \frac{\partial
\mathcal{L}_{SM}}{\partial(D_{\mu} \phi_{i})} D^{\nu}\phi_{i} + 2
\frac{\partial \mathcal{L}_{SM}}{\partial(F^{a}_{\mu\lambda})}
F^{a \nu}_{\lambda}\, .
\end{equation}

Notice that once we have fixed the normalization of the Goldstino
and SM-fields, this coupling is completely determined and
model-independent. This is the low-energy theorem for
supersymmetry (SUSY). However, it turns out that the above procedure
does not lead to the most general effective action invariant under
non-linear supersymmetry. It is possible to find additional
supersymmetric terms which could in principle be added to the above
effective Lagrangian. Since these terms would need to be supersymmetric
by themselves, their overall normalization is not determined within the
effective field theory, but depends on the underlying fundamental
theory. If we restrict ourselves to terms suppressed by at most two
powers of $\kappa$, then supersymmetry allows only a small number of
such terms. In the next section, we use the superfield formalism of
non-linear supersymmetry~\cite{Samuel:1982uh} to find the complete list.

\section{The superfield formalism}
\setcounter{equation}{0}

Starting with a generic field $\phi_{i}$ in some representation of
supersymmetry, we can systematically construct a superfield $\Phi_i$ with
$\phi_{i}$ being its lowest component~\cite{Wess:cp}:
\begin{equation}
\Phi_{i}(x, \theta, \bar{\theta}) = \exp(\theta Q + \bar{\theta}
\bar{Q}) \times \phi_{i}(x)\, ,
\label{sfield}
\end{equation}
where the multiplication symbol $\times$ means that the supercharges
$Q$ and $\bar{Q}$ operate
on $\phi_{i}$ in the appropriate representation. If we apply this
prescription to the case of a field transforming in the standard
realization, we
obtain
\begin{eqnarray}
\Phi_{i}(x, \theta, \bar{\theta}) & = & \phi_{i}(x) - i \kappa (
\lambda \sigma^{\mu} \bar{\theta} - \theta \sigma{^\mu}
\bar{\lambda} )
\partial_{\mu} \phi_{i}(x) + ... \nonumber \\ & = & \phi_{i}(x) +
\kappa \Lambda^{\mu}
\partial_{\mu} \phi_{i}(x) + ...
\end{eqnarray}
which contains ordinary derivatives of $\phi_{i}$. Therefore, if
$\phi_{i}$ transforms covariantly in some representation of the gauge group,
then this representation does not carry over ``nicely'' to the
corresponding superfield. However, it is possible to rewrite
the superfield without explicitly using derivatives. To this end, we
define a new variable
\begin{equation}
\tilde{x}^{\mu} \equiv x^{\mu} + \Lambda^{\mu}(\tilde{x}).
\end{equation}

The claim is then that, starting with some field in the standard
realization, we can obtain the corresponding superfield simply by
making the replacement $x \rightarrow \tilde{x}$. In other words
\begin{equation}
\Phi_{i}(x, \theta, \bar{\theta}) \equiv \phi_{i}(\tilde{x}).
\end{equation}
It is understood here that the right-hand side needs to be
expanded in a formal power-series around $x$. To check this claim,
we first note that $\Lambda^{\mu}(x)$ transforms in the standard
realization:
\begin{equation}
\delta \Lambda^{\mu}(x) = \kappa \Lambda^{\nu}(x) \partial_{\nu}
\Lambda^{\mu}(x)\, .
\label{lambda_transformation}
\end{equation}
Then we expand $\phi_{i}(\tilde{x})$ in a power series. A
straightforward calculation using~(\ref{field_transformation}) and~(\ref{lambda_transformation}) shows that this can be written as
\begin{eqnarray}
\phi_{i}(\tilde{x}) & = & \phi_{i}(x) + \delta \phi_{i}(x) +
\frac{1}{2!} \delta^{2}\phi_{i}(x) +
\frac{1}{3!}\delta^{3}\phi_{i}(x) + \frac{1}{4!} \delta^{4}
\phi_{i}(x) \nonumber \\ & = &\exp(\theta Q + \bar{\theta} \bar{Q}) \times
\phi_{i}(x)\, .
\end{eqnarray}
Thus, the claim follows by comparing with~(\ref{sfield}). In the same
way, it can be shown that the Goldstino superfield can be written
as
\begin{equation}
G_{\alpha}(x, \theta, \bar{\theta}) =
\frac{\theta_{\alpha}}{\kappa} + \lambda_{\alpha}(\tilde{x})\, .
\end{equation}

When expressed in this way, the superfield $\Phi$ does not involve
any explicit derivatives. Suppose $\phi_{i}(x)$ transforms in some
representation of the gauge group,
\begin{equation}
\phi'_{i}(x) = \exp(i \Omega^{A}(x) t^{A} )_{ij} \phi_{j}(x)\, .
\end{equation}
Since gauge transformations correspond to a symmetry of the
theory it is natural to require that $\phi'_{i}$ should also
transform in the standard realization (this in turn implies that
the parameter $\Omega^{A}(x)$ must transform in the standard
realization as well). Following the discussion above, the
transformation of the corresponding superfields is then
obtained by substituting $x \rightarrow \tilde{x}$:
\begin{equation}
\Phi'_{i}(x,\theta, \bar{\theta}) = \exp(i \Omega^{A}(\tilde{x})
t^{A} )_{ij} \Phi_{j}(x,\theta, \bar{\theta})\, .
\end{equation}
The superfield transforms just like its lowest component field
except that the parameter of the transformation is now itself a
superfield. Using the fact that $\tilde{x}$ is real, we also have
\begin{equation}
{\Phi'}^{\dagger}_{i}(x,\theta, \bar{\theta}) = \Phi^{\dagger}_{j}(x,\theta,
\bar{\theta})\ \exp( - i \Omega^{A}(\tilde{x}) t^{A} )_{ji}\, .
\end{equation}

This procedure works for any field in the standard realization and in
particular for SUSY-covariant derivatives of such fields as defined in~(\ref{susy_covariant_derivative}). In this way, we can construct
derivatives of superfields:
\begin{equation}
\mathcal{D}_{\mu} \Phi_{i}(x, \theta, \bar{\theta}) \equiv
((A^{-1})_{\mu}^{\ \nu} D_{\nu}\phi_{i})(\tilde{x})\, .
\label{SUSYderivative}
\end{equation}
Under gauge transformations, $\mathcal{D}_{\mu}\Phi_{i}$ transforms
just like $\Phi_{i}$. With these ingredients it is a
straightforward task to construct gauge-invariant quantities out
of superfields and their derivatives. Any function of $\Phi_{i}$,
$\Phi^{\dagger}_{i}$ and their covariant derivatives (as defined in~(\ref{SUSYderivative})) that is invariant under global gauge
transformations is also invariant under local gauge transformations.

\section{Supersymmetric Goldstino couplings}
\setcounter{equation}{0}
We shall at first consider only scalars and gauge fields. Starting
from fields in the standard realization we obtain the
corresponding superfields by substituting $x \rightarrow
\tilde{x}$. The relevant superfields are the Goldstino, the
field-strength tensor, the scalar field and possibly their
derivatives:
\begin{eqnarray}
\kappa G_{\alpha}(x,\theta, \bar{\theta}) & = & \theta_{\alpha} +
\kappa \lambda_{\alpha}(\tilde{x}) \\
\mathcal{F}^{a}_{\mu\nu}(x, \theta, \bar{\theta}) & = & (
(A^{-1})_{\mu}^{\ \sigma} (A^{-1})_{\nu}^{\ \rho} F^{a}_{\sigma
\rho} )(\tilde{x}) \\
\Phi_{i}(x, \theta, \bar{\theta}) & = & \phi_{i}(\tilde{x})
\end{eqnarray}
The field $\kappa G_{\alpha}$ has mass dimension $-\frac{1}{2}$
while $\mathcal{F}^{a}_{\mu\nu}$ and $\Phi_{i}$ have dimensions 2
and 1, respectively. Notice that when expanding the above
superfields, every insertion of the Goldstino field $\lambda_{\alpha}$
comes with exactly one power of $\kappa$.

Out of these ``elementary" superfields we can construct
gauge-invariant superfield Lagrangians $\mathcal{L}^{SF}$. The
corresponding action is obtained as usual by a superspace
integration:
\begin{equation}
\mathcal{S} = \int d^4x d^2\theta d^2{\bar{\theta}}
\mathcal{L}^{SF}.
\end{equation}
Consider specifically an operator $\mathcal{O}$ built out of the
above superfields. The corresponding supersymmetric field operator
is given by
\begin{equation}
O = \int d^{2}\theta d^{2}\bar{\theta} \mathcal{O} = O^{(0)} +
\kappa^{2} O^{(2)} + \kappa^{4} O^{(4)}+ ... \label{exp1}
\end{equation}
where the right-hand side denotes an expansion in powers of
$\kappa$. Since there is one power of $\kappa$ for each $\lambda$,
the operator $O^{(i)}$ contains $i$ Goldstino fields.
Interactions between the Goldstino, scalars, gauge fields and
an even number of matter fermions must contain an even number of
$\lambda$, so the expansion~(\ref{exp1}) contains in this case only even
powers of $\kappa$. However, as we will see below, there are also
interactions involving an odd number of Goldstinos, and thus, odd powers
of $\kappa$.

If $\mathcal{O}$ has mass dimension $D \geq 0$ then $O$ has
dimension $d = D + 2$ and $O^{(i)}$ has dimension
$d_{i}=D+2i+2.$\footnote{We shall always use capital $D$ to denote
the mass dimension of superfields and lower-case $d$ for the
dimension of ordinary fields.} Thus, on dimensional grounds, the
operator $O$ gives an effective Lagrangian term of the form
\begin{equation}
\mathcal{L}_{O} = g \kappa^{\frac{D}{2}-1} O = g
\kappa^{\frac{D}{2}-1} O^{(0)} + g \kappa^{\frac{D}{2}+1} O^{(2)}
+ g \kappa^{\frac{D}{2}+3} O^{(4)} +...
\label{exp2}
\end{equation}
where g is some numerical constant. We shall only be interested in
low-dimensional interactions suppressed by at most two powers of
$\kappa$. In that case it is easy to see that it is enough to consider
superfields $\mathcal{O}$ with $D \leq 2$ and we only need to keep
$O^{(0)}$ and $O^{(2)}$ in the expansion (\ref{exp2}). Of course
$O^{(0)}$ does not involve any Goldstinos and exists already in the
effective action independently of non-linear supersymmetry, while
$O^{(2)}$ for $D>2$ has dimension bigger than 8 and we drop.

To find all possible couplings of the Goldstino to Standard Model
fields consistent with non-linear supersymmetry, we shall proceed
in two ways. Usually, it will be simpler to write down all gauge
invariant superfields $\mathcal{O}$ with $D \leq 2$ and then
perform the superspace integration to obtain the corresponding
contribution to the effective action. However, in a few cases it will
be more convenient to work in the other way around: first find
all viable candidates for $O^{(2)}$ and then show that these can
indeed be realized in terms of superfields. This will be trivial whenever
all Goldstino fields are acted upon by derivatives. In this case, we
simply have to replace all fields by the corresponding superfields and
multiply the whole expression by $\kappa^4 G^2 \bar{G}^2$.  This is then
manifestly supersymmetric and after carrying out the integration over the
fermionic coordinates, we recover $O^{(2)}$ as the first term in the
expansion in powers of $\kappa$.  In fact this is equivalent to the
``supersymmetrization'' procedure introduced in ref.~\cite{Clark:1996aw}.

\subsection{Couplings to gauge fields} \label{couplings_to_gauge_fields}
We first assume for simplicity that there is no $U(1)$ factor in
the gauge group. In that case, gauge invariance requires that the
field-strength enters $\mathcal{L}^{SF}$ quadratically, which brings
already a factor of dimension 4. Because the
Goldstino superfields are Grassmann variables, any superfield can
contain at most four of them if no derivatives are involved. Since they
have dimension $-\frac{1}{2}$, no terms of dimension $D<2$ are possible
while $D=2$ operators necessarily involve four Goldstino superfields.
 From the identities
\begin{equation}
\begin{gathered}
G_{\alpha} G_{\beta} = \frac{1}{2} G^{2} \epsilon_{\alpha \beta}
\\
\bar{G}_{\dot{\alpha}} \bar{G}_{\dot{\beta}} = - \frac{1}{2}
\bar{G}^{2} \epsilon_{\dot{\alpha} \dot{\beta}} \label{id1}
\end{gathered}
\end{equation}
it follows that there is only one possibility:
\begin{equation}
\mathcal{O}_{1} = -\frac{1}{4} \kappa^4 \mathcal{F}^{a}_{\mu\nu}
\mathcal{F}^{a \mu \nu} G^{2}\bar{G}^{2}\, .
\end{equation}

If $\mathcal{S}(x,\theta, \bar{\theta}) \equiv s(\tilde{x})$ is a
superfield with $s(x)$ its lowest component, then:
\begin{equation} (\kappa^4 \mathcal{S} G^{2}
\bar{G}^{2})_{\theta^2 \bar{\theta^2}} = s(x) + i \kappa^{2}
\lambda \sigma^{\mu} \overset{\leftrightarrow}{\partial}_{\mu}
\bar{\lambda} s(x) + O(\kappa^{4})\, .
\label{id2}
\end{equation}
With the help of this identity and the definition~(\ref{newF}) we
obtain
\begin{equation}
O_{1} = -\frac{1}{4} F^{a}_{\mu\nu} F^{a \mu \nu} + i \kappa^2 (
\lambda \overset{\leftrightarrow}{\partial_{\mu}} \sigma^{\nu}
\bar{\lambda}) (-F^{a}_{\nu\sigma} F^{a \sigma\mu} -
\frac{\delta^{\mu}_{\ \nu}}{4} F^{a}_{\alpha\beta} F^{a \alpha
\beta}) + ...
\end{equation}
The quantity in parenthesis in the second term is just the
contribution of the field-strength to the energy-momentum tensor
and so $O_{1}$ simply reproduces the coupling required by the
low-energy theorem.

In principle we could also have a term
\begin{equation}
\mathcal{O}_{2} = - \frac{\theta}{64 \pi^{2} } \kappa^4
\epsilon^{\mu\nu\sigma\rho} \mathcal{F}^{a}_{\mu\nu}
\mathcal{F}^{a}_{\sigma \rho} G^{2}\bar{G}^{2}\, .
\end{equation}
This would again reproduce the result of the low-energy theorem,
but since $\Tr(F \tilde{F})$ is a total derivative, its contribution to
the energy-momentum tensor vanishes.

If the gauge group contains $U(1)$ factors, then we should also
consider terms linear in $F_{\mu\nu}$. Using~(\ref{id1}),
it is easy to see that no such terms exist at dimension $D=0$. To
construct $D=1$ operators we need to use derivatives of the
Goldstino superfield. If we start with terms containing a single
superderivative,
then there are two possible operators:
\begin{eqnarray}
\mathcal{O}_{3} & = & \kappa^{4} \mathcal{F}_{\mu\nu}G^{2}\
\mathcal{D}^{\mu}G \sigma^{\nu} \bar{G} \nonumber \\
\mathcal{O}_{4} & = & \kappa^{4} \mathcal{F}_{\mu\nu} \epsilon^{\mu
\nu \alpha \beta} G^{2}\
\mathcal{D}_{\alpha}G \sigma_{\beta} \bar{G}
\end{eqnarray}
If we take $\mathcal{S}_{\mu\nu}(x, \theta, \bar{\theta})$ to be
some superfield with lowest component $s_{\mu\nu}(x)$, then up to
total derivatives we have the identity

\begin{equation}
(\kappa^{4} \mathcal{S}_{\mu\nu} G^{2} \mathcal{D}^{\mu}G
\sigma^{\nu} \bar{G})_{\theta^{2} \bar{\theta}^{2}} =
\frac{i}{2} \kappa^{2} s_{\mu\nu}\ \partial_{\alpha}\lambda
\sigma^{\alpha} \bar{\sigma}^{\nu} \mathcal{D}^{\mu}\lambda +
O(\kappa^{4})\, . \label{id3}
\end{equation}
This is proportional to the equations of motion and therefore
$\mathcal{O}_{3}$ and $\mathcal{O}_{4}$ do not contribute at order
$\kappa^{2}$.

If we add an extra derivative then there are additional possibilities.
These are obtained by considering all possible contractions of indices in
the following superfield:
\begin{equation}
\mathcal{O}_{5} = \kappa^4  G^2 \bar{G}^2 \mathcal{F}_{\mu\nu}
\mathcal{D}_\alpha G J^{\beta \gamma}_{(\frac{1}{2},0)}
\mathcal{D}_{\delta}G\, ,
\end{equation}
where $J^{\mu\nu}_{(\frac{1}{2},0)}$ are the generators of the
$(\frac{1}{2},0)$ representation of the Lorentz group:
\begin{equation}
J^{\mu\nu}_{(\frac{1}{2},0)} = \frac{i}{4}(\sigma^{\mu}
\bar{\sigma}^{\nu} - \sigma^{\nu} \bar{\sigma}^{\mu} )\, .
\end{equation}
Carrying out the fermionic integration and using the equations of motion
for the Goldstino it is easy to see that these terms again do not
contribute at order $\kappa^2$. This exhausts all possibilities for $D=1$.

We still need to check the case $D=2$. Instead of working with
superfields it turns out to be more convenient to find all
dimension $d=8$ field operators containing two Goldstinos and being
linear in $F_{\mu\nu}$. The prototype for these operators is
$F_{\mu\nu}\partial_{\alpha}\lambda \sigma_{\beta}
\partial_{\gamma} \partial_{\delta} \bar{\lambda}$. Looking for all
possible ways of contracting the indices with the metric or
$\epsilon^{\mu\nu\sigma\rho}$, it is easy to show that all these
terms except one either vanish or are proportional to the equations of
motion. The remaining term is\footnote{The analogous term containing
$\tilde{F}_{\mu\nu}=\frac{1}{2} \epsilon_{\mu\nu\sigma\rho}
F^{\sigma\rho}$ brings nothing new since up to terms proportional to the
equations of motion we have $(\partial^{\alpha}\lambda \sigma^{\mu}
\partial^{\nu}\bar{\lambda})\partial_{\alpha} \tilde{F}_{\mu\nu} =
-i(\partial^{\alpha}\lambda \sigma^{\mu}
\partial^{\nu}\bar{\lambda})\partial_{\alpha}F_{\mu\nu}$.}
\begin{equation}
O^{(2)}_{5} = \kappa^2 \partial^{\alpha} \lambda \sigma^{\mu}
\partial^{\nu} \bar{\lambda} \partial_{\alpha}F_{\mu\nu}\, .
\end{equation}
Since both Goldstino fields are coming with derivatives, it immediately
follows from the discussion at the end of last subsection that this term
is supersymmetric. It was already found in ref.~\cite{Clark:1998aa}.

\subsection{Couplings to scalar fields}
We shall proceed systematically beginning with interactions which
are quartic in the scalar fields.
\begin{enumerate}
\item \textbf{Quartic interactions} 

No $D=0,1$ terms are possible since these would require
respectively 8 and 6 Goldstino superfields. There is a single
$D=2$ operator:
\begin{equation}
\mathcal{O} = \frac{\kappa^{4}}{4!} M_{ijkl}
\Phi^{(1)}_i\Phi^{(2)}_j\Phi^{(3)}_k\Phi^{(4)}_l G^{2} \bar{G}^{2}\, .
\label{cubic}
\end{equation}
The superscripts on the scalar fields indicate that they might transform
in  different representations of the gauge group which are coupled
through the  Clebsch-Gordan coefficients $M_{ijkl}$. Using the identity
(\ref{id2}), this results in a $\phi^{4}$ term together with the coupling
of the Goldstino to the corresponding contribution to the energy-momentum
tensor. In fact this coupling is proportional to the equations of motion
and can be omitted.

\item \textbf{Cubic interactions} 

Again there can be no $D=0$ terms. For $D=1$, we can have a term
similar to~(\ref{cubic}) but cubic in $\Phi$. This again
reproduces the low-energy coupling to the stress-energy tensor and in any
case it can be omitted since it is proportional to the equations of
motion. For $D=2$, restricting first to couplings containing at most one
superderivative, we find two terms:
\begin{eqnarray}
\mathcal{O}_{1} & = & \frac{\kappa^{2}}{3!}
\Phi^{(1)}_{i}\Phi^{(2)}_{j}\Phi^{(3)}_{k} M_{ijk}\ G^2 \\
\mathcal{O}_{2} & = & \frac{\kappa^{4}}{3!}
\Phi^{(1)}_{i}\Phi^{(2)}_{j}\Phi^{(3)}_{k} M_{ijk}G^{2}\
\mathcal{D}_{\mu}G \sigma^{\mu} \bar{G}
\end{eqnarray}
Since $\mathcal{D}_{\mu}(\kappa G_{\alpha})= (\kappa
\partial_{\mu} \lambda_{\alpha})(\tilde{x}) + O(\kappa^{3})$, the
operator $\mathcal{O}_{2}$, up to higher order terms, is proportional
to the equations
of motion and can be omitted. $\mathcal{O}_{1}$ on the
other hand gives something new. Carrying out the superspace
integration, we obtain the supersymmetric action:
\begin{equation}
S = i \kappa^{2} \int d^4x \phi^{(1)}_{i}\phi^{(2)}_{j}\phi^{(3)}_{k} M_{ijk}\
(\partial_{\mu} \lambda J^{\mu\nu}_{(\frac{1}{2},0)}
\partial_{\nu} \lambda)\, .
\label{cubic_action}
\end{equation}
Adding a second superderivative yields additional terms but all of these
reproduce (\ref{cubic_action}). Terms with more than two superderivatives
generate interactions with $d>8$, which we omit.

\item \textbf{Quadratic interactions} 

Here again it is more convenient to work directly with ordinary fields and first look for possible candidates for
$O^{(2)}$. These operators should contain two Goldstinos
together with some derivatives, one of which at least must act on
the Goldstinos since the coupling must vanish for constant
$\lambda$. At mass dimension $d=6$ we find
\begin{equation}
O^{(2)}_{1} = i \lambda \sigma^{\mu}
\overset{\leftrightarrow}{\partial_{\mu}} \bar{\lambda} \phi^{(1)}_{i}
\phi^{(2)}_{j} M_{ij}\, ,
\end{equation}
which is proportional to the equations of motion. At $d=7$ we have four
possibilities:
\begin{eqnarray*}
O_{1}^{(2)} & = & (D_{\mu}\phi^{(1)}_{i} \phi^{(2)}_{j} M_{ij}-
\phi^{(1)}_{i} D_{\mu}\phi^{(2)}_{j} M_{ij} ) \lambda
\partial^{\mu} \lambda \\
O_{2}^{(2)} & = & (D_{\mu}\phi^{(1)}_{i} \phi^{(2)}_{j} M_{ij}-
\phi^{(1)}_{i} D_{\mu}\phi^{(2)}_{j} M_{ij} ) \lambda
J^{\mu\nu}_{(\frac{1}{2},0)}
\partial_{\nu} \lambda \\
O_{3}^{(2)} & = & \phi^{(1)}_{i} \phi^{(2)}_{j} M_{ij}
\partial_{\alpha}\lambda \partial^{\alpha} \lambda \\
O_{4}^{(2)} & = & \phi^{(1)}_{i} \phi^{(2)}_{j} M_{ij}\ \partial_{\mu}
\lambda J^{\mu\nu}_{(\frac{1}{2},0)}
\partial_{\nu} \lambda
\end{eqnarray*}

Up to total derivatives, $O_{1}^{(2)}$ can be rewritten as
\begin{equation}
\frac{1}{2} (D^{2}\phi_{i}\phi_{j} M_{ij} - \phi_{i} D^{2}\phi_{j}
M_{ij})\lambda^{2}\, ,
\end{equation}
which is proportional to the equations of motion of the scalar
fields. Also, up to terms proportional to the equations of motion
of the Goldstino, $O^{(2)}_1$ and $O^{(2)}_2$ are equivalent. The same
is true for $O^{(2)}_3$ and $O^{(2)}_4$. This leaves us with only
$O_{4}^{(2)}$ which is trivially supersymmetric. According to~(\ref{exp2}) with $D=1$, this operator then contributes to the action a
term
\begin{equation}
S = \kappa^{\frac{3}{2}} \int d^{4}x \phi^{(1)}_{i} \phi^{(2)}_{j}
M_{ij} \partial_{\mu}\lambda J^{\mu\nu}_{(\frac{1}{2},0)}
\partial_{\nu}\lambda\, .
\end{equation}

Finally, we need to consider $d=8$ operators. There are three
possibilities
\begin{eqnarray*}
O_{1}^{(2)} & = & (D_{\mu}\phi^{(1)}_{i} \phi^{(2)}_{j} M_{ij}- \phi^{(1)}_{i}
D_{\mu}\phi^{(2)}_{j} M_{ij} )
\partial_{\alpha} \lambda \sigma^{\mu} \partial^{\alpha} \label{O2quadratic}
\bar{\lambda} \\
O_{2}^{(2)} & = & (D_{\mu}\phi^{(1)}_{i} \phi^{(2)}_{j} M_{ij}- \phi^{(1)}_{i}
D_{\mu}\phi^{(2)}_{j} M_{ij} ) \epsilon^{\mu\nu\alpha\beta}
\partial_{\nu} \lambda \sigma_{\alpha}
\partial_{\beta}\bar{\lambda} \\
O_{3}^{(2)} & = & D_{\alpha}\phi^{(1)}_{i} D_{\beta} \phi^{(2)}_{j}
M_{ij} \lambda
\sigma^{\alpha} \overset{\leftrightarrow}{\partial^\beta}
\bar{\lambda}
\end{eqnarray*}
Since $D_{\alpha}\phi^{(1)}_{i} D_{\beta} \phi^{(2)}_{j} M_{ij}$ is the
contribution to the energy-momentum tensor coming from the kinetic term
of the scalar fields, $O_{3}^{(2)}$ corresponds to the interaction of the
low-energy theorem. On the other hand, $O_{1}^{(2)}$ and $O_{2}^{(2)}$ are
equivalent due to the on-shell identity
\begin{equation}
\partial_{\alpha} \lambda \sigma^{\mu} \partial^{\alpha}
\bar{\lambda}= -i \epsilon^{\mu\nu\alpha\beta}\partial_{\nu}
\lambda \sigma_{\alpha} \partial_{\beta}\bar{\lambda}\, .
\end{equation}

Thus, we are left with the operator $O_{1}^{(2)}$, which is
supersymmetric and whose contribution to the action is given by~(\ref{exp2}):
\begin{equation}
S = \kappa^{2} \int d^{4}x\ (D_{\mu}\phi^{(1)}_{i} \phi^{(2)}_{j}
M_{ij}- \phi^{(1)}_{i} D_{\mu}\phi^{(2)}_{j} M_{ij} )
\partial_{\alpha} \lambda \sigma^{\mu}
\partial^{\alpha}\bar{\lambda}. \label{SSLL}
\end{equation}

This exhausts all possible couplings to scalar fields.
\end{enumerate}

\subsection{Couplings to scalars and gauge field-strengths}

Operators with two scalars should contain for dimensional reasons only
one power of $\mathcal{F}^a_{\mu\nu}$ together with four Goldstino
superfields. From the identity~(\ref{id1}) it then follows that the Goldstinos must appear in the form $G^2 \bar{G}^2$ and this implies that the spacetime indices of $\mathcal{F}^a_{\mu\nu}$ must be contracted among
themselves, giving a vanishing result. We conclude that these couplings can only contain a single scalar field which must be in the adjoint representation.
Generically they have the form:
\begin{equation}
\mathcal{O} = \Phi^a \mathcal{O'}^{a}(\mathcal{F}^a_{\mu\nu}, G_{\alpha},
\bar{G}_{\beta})\, ,
\end{equation}
where $\mathcal{O'}$ has dimension $D \leq 1$. From the analysis of
Section~\ref{couplings_to_gauge_fields}, it then follows
that none of these couplings can contribute at order $\kappa^2$.

\subsection{Couplings to fermions} \label{couplings_to_fermions}

If we assume that lepton and baryon number are conserved, any such
interaction must involve an equal number of matter fermions and
anti-fermions. This case has already been considered in
ref.~\cite{Brignole:1997pe} and here we just quote the result. It was
found that up to order
$\kappa^{2}$ there is a single possible term besides the standard coupling to
the energy-momentum tensor:
\begin{equation}
S = \kappa^{2} \int d^{4}x (f \partial^{\mu} \lambda)(\bar{f}
\partial_{\mu} \bar{\lambda})\, .
\end{equation}

However, since lepton and baryon number conservations result from
accidental global symmetries of the Standard Model that may be broken by
non-renormalizable interactions, it is natural to consider also terms
which do not preserve these symmetries. The above four-fermion coupling
then generalizes to
\begin{equation}
S_1 = \kappa^{2} \int d^{4}x M_{ij}\ (f^{(1)}_i \partial^{\mu}
\lambda)(\bar{f}^{(2)}_j \partial_{\mu} \bar{\lambda})\, .
\end{equation}
Furthermore, there is an additional four-fermion interaction\footnote{The action $S_3 = \int d^4 x M_{ij} (f_i^{(1)} \partial_\mu \lambda)(f_j^{(2)} \partial^\mu \lambda)$, which is also supersymmetric, is related to~(\ref{new_fermion_interaction}) by a Fierz rearrangement. We thank A. Brignole for pointing this out to us.}:
\begin{equation}
S_2  =  \int d^4 x M_{ij} (f^{(1)}_i
f^{(2)}_j)(\partial_{\mu}\lambda \partial^{\mu}\lambda)  \label{new_fermion_interaction}
\end{equation}
We can also consider terms which are linear in $\lambda$, allowing for
several dimension 6 operators:
\begin{eqnarray}
O_{1} & = & \kappa (f^{a} \sigma_{\alpha} \partial_{\beta}
\bar{\lambda}) F^{a}_{\mu\nu} \epsilon^{\alpha \beta\mu\nu} \nonumber \\
O_{2} & = & \kappa (f^{a} \sigma^{\alpha} \partial^{\beta}
\bar{\lambda}) F^{a}_{\alpha\beta} \nonumber \\
O_{3} & = & \kappa M_{ij}(f^{i} \partial_{\alpha} \lambda)
D^{\alpha}\phi^{j}
\end{eqnarray}

All these operators are supersymmetric. Actually, using the equations of
motion for $\lambda$, $O_1$ is proportional to $O_2$. Moreover, $O_{2}$
cannot appear in the Standard Model because it requires fermions
transforming in the adjoint representation. One could however use a
gauge singlet fermion, such as a right-handed neutrino, coupled to the
hypercharge field-strength. On the other hand, $O_{3}$ is allowed
provided that $f^i$ is a lepton doublet and $\phi^j$ is the Higgs field.
There can be additional terms linear in $\lambda$ of order
$\kappa^{2}$ but we shall not attempt to construct these systematically
here, since they are too many.

This exhausts all possibilities for coupling the Goldstino to
Standard Model fields consistent with non-linear supersymmetry.
The new couplings we found are of the same, or even lower, order in
$\kappa$ as the coupling to the energy-momentum tensor and must
therefore be taken into account in the low-energy theory. Their
(dimensionless) coefficients are model-dependent and can be determined
by the underlying fundamental theory. In the next section we shall
determine these values in string theory with D-branes. For future
reference, we summarize the full list of operators in Appendix A.

Before proceeding, there is an important technical point that needs to
be mentioned. In this section, we have found two interactions of
order $\kappa$: (\ref{O1}) and (\ref{O2}). They both give rise to
three-point functions involving massless particles, which obviously
vanish on-shell because of the derivatives. Since in string theory we
can only do on-shell calculations, these operators cannot be determined
``directly'' by computing three-particle interaction amplitudes. A way
to compute the coefficients $C_{1}$ and $C_2$ is to look instead at
four-point tree-level amplitudes, obtained by combining two order
$\kappa$ vertices. These are in fact contact terms, because the
propagator of the intermediate massless state gets canceled.
Furthermore, these reducible contact terms must be (non-linear)
supersymmetric, because they arise by combining two supersymmetric
vertices. Thus, they are indistinguishable from other irreducible
contributions to the amplitudes, obtained by order $\kappa^2$ terms in
the list of operators. This makes it difficult to disentangle the two
contributions.

As we shall see, it is possible to obtain irreducible amplitudes by
considering generic string interactions on intersecting stacks of
D-branes where the intermediate states generating reducible
contributions are massive. Assuming that all effective couplings depend
continuously on the intersection angles, it is then possible to obtain
their values also for single stacks of D-branes.

A similar problem occurs with the 3-point interaction~(\ref{O7}).
However, in this case the contact term generated at four-point level is
of order $\kappa^4$ which we do not compute. Thus, we do not determine
in this work the coefficient $C_6$. We also leave undetermined the
coefficient $C_8$ which requires the computation of a 5-point function.

In Sections \ref{single_stack} and \ref{intersecting_stacks}, we compute
all the relevant string amplitudes. In  Section \ref{effective_action},
we combine the results to extract the values of the coefficients $C_i$
and of the Goldstino decay constant $\kappa$.

\section{Single stack of Dp-branes} \label{single_stack}
\setcounter{equation}{0}
Our fist task is to identify the Goldstino in the spectrum.
We consider spacetime to be the product of (3+1)-dimensional Minkowski
space $M_{4}$ with some internal six-dimensional compact manifold $M$
which we assume to preserve all 32 supersymmetries of the type II
superstring theory. We shall first consider a single stack of $N$
D$p$-branes with $p \geq 3$. We identify the first 4 dimensions of
D-brane world-volume with $M_{4}$ while the remaining $p-3$ dimensions
are wrapped around a cycle of the internal manifold.
The presence of branes breaks half of the supercharges spontaneously and
from the point of view of (3+1)-dimensions we have ${\cal N}=4$
supersymmetry. The massless open string spectrum consists of an
${\cal N}=4$ vector multiplet transforming in the adjoint representation
of $U(N)$~\footnote{In the general case, one could add orientifolds and
break (part of) the supersymmetry, but their presence does not modify our
analysis.}. For each of the four broken supersymmetries there is a pair of
massless Goldstone fermions (Goldstinos) forming a two-component Weyl
spinor in four dimensions. These carry the same quantum numbers as the
broken supercharges. Since the supercharges commute with the generators
of the gauge group, the Goldstinos should be gauge singlets and must
therefore be identified with the gauginos in the
$U(1)$ vector mutliplet which is part of $U(N)$. We shall pick one of
these Goldstinos and its CPT conjugate and calculate its effective
interactions with the particles in the $SU(N)$ vector multiplet. This
will allow to extract an expression for the supersymmetry breaking scale
$\kappa$ in terms of the string scale $M_{s}$ and to determine the
coefficients of the (model-dependent) supersymmetric operators obtained
in the previous section.

\subsection{Interactions with gauge bosons}

As is well-known, the tree-level interaction amplitude involving open
string states is obtained by evaluating correlation functions of vertex
operators on the boundary of the disc. By a suitable conformal
transformation, the disc can be mapped onto the upper half-plane with the
vertex operators located on the real axis. After integrating over the
positions of the vertices we must divide by the volume of
$PSL(2,R)$, the conformal group on the disc. Alternatively, we can
fix the location of three vertex operators to some arbitrary
positions which we shall choose to be $x=0,1,\infty$. Using the
conformal symmetry of the disc, the range of integration of the
fourth operator can be mapped to the interval $[0,1]$. The
amplitude has then the form
\begin{equation}
\begin{split}
\mathcal{A}(1,2,3,4) = A(1,2,3,4)\
\Tr(\lambda^{1}\lambda^{2}\lambda^{3}\lambda^{4} +
\lambda^{4}\lambda^{3}\lambda^{2}\lambda^{1}) \\
  +  A(4,2,3,1)\
\Tr(\lambda^{4}\lambda^{2}\lambda^{3}\lambda^{1} +
\lambda^{1}\lambda^{3}\lambda^{2}\lambda^{4}) \\ +  A(1,2,4,3)\
\Tr(\lambda^{1}\lambda^{2}\lambda^{4}\lambda^{3} +
\lambda^{3}\lambda^{4}\lambda^{2}\lambda^{1})
\label{totamplitude}
\end{split}
\end{equation}
with
\begin{equation}
A(1,2,3,4) = i C_{D}\ \int_{0}^{1} dx
<(c\nu_{q_{1}})(0)\nu_{q_{2}}(x) (c\nu_{q_{3}})(1)
(c\nu_{q_{4}})(\infty)>\, .
\label{amplitude}
\end{equation}
Here $\lambda^{i}$ is the Chan-Paton matrix that comes with the
vertex operator $\nu_{q_{i}}(x)$. The subscripts $q_{i}$ denote the ghost
numbers which for the case of the disc must add up to $-2$ in order to
cancel the superconformal background charge~\cite{Friedan:1985ge}. The
constant $C_{D}$ depends only on the topology of the world-sheet
and can be determined by unitarity. It is given by
\begin{equation}
C_{D} = \frac{1}{2 {\alpha'}^{2} g_{YM}^{2} V_{c}}\, ,
\label{DiscConstant}
\end{equation}
where $g_{YM}$ is the four-dimensional Yang-Mills coupling\footnote{Here
we use a somewhat non-standard convention where the generators of the
gauge group are normalized according to $\Tr(T^{a}T^{b}) = \delta_{ab}$.
This implies in particular that the canonically normalized kinetic term
for the gauge bosons is given by $-\frac{1}{4}\Tr(F_{\mu\nu}F^{\mu\nu})$.}
and
$V_{c}$ is the compactification volume along the D$p$-brane world-volume.
Finally, the gauge-fixing procedure requires that each ``fixed'' vertex
operator is accompanied by a c-ghost insertion.

It will be convenient to express all amplitudes in terms of the
Mandelstam variables:
\begin{equation}
s= -(k_{1} + k_{2})^{2} \qquad t = -(k_{1} + k_{3})^{2} \qquad u =
-(k_{1} + k_{4})^{2}\, ,
\end{equation}
where $k_i$ is the spacetime momentum of the state $i$.

We shall first evaluate the interaction of two Goldstinos and two
gauge bosons. As stated above, the spectrum contains eight
Goldstinos characterized by their helicities in the internal directions.
We shall conventionally choose the Goldstino
to be the abelian gaugino with internal helicities given by
$\{+\ +\ +\}$, which by virtue of the GSO projection will have
positive spacetime chirality. Its CPT conjugate will then have internal
helicities
$\{-\ -\ -\}$ and negative spacetime chirality. The corresponding vertex
operators in the $(-\frac{1}{2})$-ghost picture are
\begin{eqnarray}
\nu^{G}_{-\frac{1}{2}}(x, k, u_{L}) & = &  (4
{\alpha'}^{3})^{\frac{1}{4}} g_{YM}\ u_{L\alpha} \Theta^{\alpha}\
e^{-\frac{\phi}{2}}e^{ - \frac{i}{2}(H_{1} +
   H_{2} + H_{3})}\ e^{ikX} \nonumber \\
\bar{\nu}^{G}_{-\frac{1}{2}}(x, k, u_{R}) & = & (4
{\alpha'}^{3})^{\frac{1}{4}} g_{YM}\ u_{R\alpha}  \Theta^{\alpha} \
e^{-\frac{\phi}{2}} e^{ \frac{i}{2}(H_{1} +
   H_{2} + H_{3})}\ e^{ikX}
\label{GoldVertex}
\end{eqnarray}
where $\phi$ is the bosonized superconformal ghost and $H_{i}$ are the
usual bosonized fermionic coordinates in the internal and transverse
directions. The $\Theta^{\alpha}$ are four-dimensional spin fields and
$u_{L,R}$ are left- and right-handed Weyl spinors (in four-component
notation), respectively. For gauge bosons $(B^\mu)$, we shall need the
vertex operators in both the $(0)$- and
$(-1)$-ghost picture:
\begin{eqnarray}
\nu_{-1}^B(x,k,\epsilon) & = & \sqrt{2\alpha'} g_{YM}\
e^{-\phi}(\epsilon
\psi)\ e^{ikX} \nonumber \\
\nu_{0}^B(x, k, \epsilon) & = & 2g_{YM}\ (\ i (\epsilon \partial X) +
\alpha' (k\psi)(\epsilon \psi)\ )\ e^{ikX} \label{GBVertex}
\end{eqnarray}
where $\epsilon$ is the polarization vector. The normalizations of
the vertices can be obtained by computing three-point
functions in the point-particle limit and comparing with the
corresponding result in Yang-Mills theory.

The correlation function we need to evaluate is:\footnote{The c-ghost
insertions are implicit here. They will always contribute a factor
$x_{13}x_{14}x_{34}$ to the correlator.}
\begin{equation}
<\nu^{G}_{-\frac{1}{2}}(x_{1}, k_{1}, u_{L1})\
\bar{\nu}^{G}_{-\frac{1}{2}}(x_{2}, k_{2}, u_{R2})\
\nu_{-1}^B(x_{3}, k_{3},\epsilon_{3})\ \nu_{0}^B(x_{4}, k_{4},
\epsilon_{4})>\, .
\end{equation}
Inserting the vertex operators, this factorizes into a product of
well-known correlators~\cite{Friedan:1985ge} and the result is
\begin{equation}
\begin{split}
<\nu_{-\frac{1}{2}}^G\ \bar{\nu}_{-\frac{1}{2}}^G \nu_{-1}^B \nu_{0}^B> =
4 \sqrt{2} {\alpha'}^{3} g_{YM}^{4} \frac{x_{13} x_{14}
x_{34}}{x_{12} x_{13} x_{23}} (2\pi)^{4} \delta^{(4)}(\sum_{i}
k_{i}) V_{c} \prod_{i<j} x_{ij}^{2\alpha'k_{i} k_{j}} \\
\biggl(     \frac{1}{\sqrt{2}} u^{T}_{L1}C\slash{\epsilon_{3}}u_{R2}
\biggl( (k_{1} \epsilon_{4}) \frac{x_{13}}{x_{14} x_{34}} + (k_{2}
\epsilon_{4}) \frac{x_{23}}{x_{24} x_{34}} \biggr)  + \\
\frac{1}{\sqrt{2}} u^{T}_{L1}C \slash{\epsilon_{4}}u_{R2}
(\epsilon_{3} k_{4}) \frac{x_{13}}{x_{34} x_{14}}  -
\frac{1}{\sqrt{2}} (\epsilon_{3}
\epsilon_{4}) u^{T}_{L1}C \slash{k_{4}}u_{R2} \frac{x_{13}}{x_{34} x_{14}} \\
  + \frac{1}{2 \sqrt{2}} u^{T}_{L1}C\slash{\epsilon_{3}}
\slash{\epsilon_{4}} \slash{k_{4}}u_{R2} \frac{x_{12}}{x_{34} x_{14}}
  \biggr) \nonumber
\end{split}
\end{equation}
where $x_{ij}=x_i - x_j$ and $C$ is the charge-conjugation matrix.
The volume factor $V_{c}$ in the numerator comes from the correlator of
exponentials:
\begin{equation}
<\prod_i e^{i k_i X}> = (2\pi)^{4} \delta^4(\sum_i k_i) V_{c}
\prod_{i<j} x^{\alpha' k_i k_j}_{ij}\, .
\label{correlator_exp}
\end{equation}
The integration over the zero modes of $X^{0,1,2,3}$ gives the delta
function, while integration over the zero modes of $X^{4,..,p}$ gives the
compactification volume along the D-brane.

To obtain the amplitude $A(1,2,3,4)$, we set $(x_{1}, x_{2}, x_{3},
x_{4}) \rightarrow (0,x,1,\infty)$ and then perform the
integration in (\ref{amplitude}). The other two permutations in~(\ref{totamplitude}) are obtained by simply permuting the
positions of the vertex operators. For example $A(4,2,3,1)$
corresponds to setting $(x_{1}, x_{2}, x_{3}, x_{4}) \rightarrow
(\infty,x, 1, 0)$ before performing the integration. Putting
everything together, we obtain the total amplitude:
\begin{equation}
\begin{split}
\mathcal{A}(\lambda_{L} \lambda_{R} \rightarrow B B) = 2 i
{\alpha'}^{2} g_{YM}^{2} & (2\pi)^{4} \delta^{(4)}(\sum_{i} k_{i})
K_{GB}(1,2,3,4) \\
& \biggl(B(s,u)\ \Tr(\lambda^{1}\lambda^{2}\lambda^{3}\lambda^{4} +
\lambda^{4}\lambda^{3}\lambda^{2}\lambda^{1}) \\ &+ B(t,u)\
\Tr(\lambda^{1}\lambda^{3}\lambda^{2}\lambda^{4} +
\lambda^{4}\lambda^{2}\lambda^{3}\lambda^{1}) \\ &+ B(s,t)\
\Tr(\lambda^{1}\lambda^{2}\lambda^{4}\lambda^{3} +
\lambda^{3}\lambda^{4}\lambda^{2}\lambda^{1}) \biggr)
\label{AALL}
\end{split}
\end{equation}
where $K_{GB}(1,2,3,4)$ is the kinematic factor
\begin{eqnarray}
K_{GB}(1,2,3,4) & = &
-u u^{T}_{L1}C \slash{\epsilon_{3}} u_{R2} (k_{2} \epsilon_{4}) -
\frac{s}{2} u^{T}_{L1}C\slash{\epsilon_{3}}\slash{\epsilon_{4}}
\slash{k_{4}}u_{R2} \nonumber
 \\ & + & t \biggl( u^{T}_{L1}C\slash{\epsilon_{3}}u_{R2}
(k_{1}\epsilon_{4}) + u^{T}_{L1}C\slash{\epsilon_{4}}u_{R2}
(\epsilon_{3} k_{4}) \nonumber \\ & - &
u^{T}_{L1} C \slash{k_{4}} u_{R2} (\epsilon_{3} \epsilon_{4}) \biggr)
\label{AA_kinematic_factor}
\end{eqnarray}
and $B(x,y)$ is given in terms of gamma functions as
\begin{equation}
B(x,y) = \frac{\Gamma(-\alpha'x) \Gamma(-\alpha'y)}{\Gamma(1-
\alpha'x- \alpha'y)} = \frac{1}{{\alpha'}^{2}xy}(1-
\frac{\pi^{2}}{6} {\alpha'}^{2}xy + \ldots)\, .
\label{Bexp}
\end{equation}

One can easily check that the amplitude~(\ref{AALL}) reproduces
correctly, in the low-energy limit, the corresponding quantum field
theory (QFT) result. Taking $\alpha' \rightarrow 0$ and using the
expansion~(\ref{Bexp}) in~(\ref{AALL}) we obtain for the $s$-channel pole
\begin{equation}
\begin{split}
\mathcal{A}^{(0)}(\lambda_{L} \lambda_{R} \rightarrow B B) \rightarrow -2i
g_{YM}^{2} (2\pi)^{4} \delta^{(4)}(\sum_{i}k_{i}) \frac{1}{s}
\Tr([\lambda^{1}, \lambda^{2}][\lambda^{3}, \lambda^{4}]) \\
\biggl( u^T_{L1} C \slash{\epsilon_{4}} u_{R2} (k_{4} \epsilon_{3}) -
u^T_{L1} C \slash{\epsilon_{3}} u_{R2} (k_{3} \epsilon_{4}) -
u^T_{L1} C \slash{k_{4}} u_{R2} (\epsilon_{3} \epsilon_{4}) \biggr)\, .
\end{split} \nonumber
\end{equation}
This agrees with the $s$-channel contribution of the corresponding QFT
amplitude with fermions in the adjoint representation of $U(N)$. The
Chan-Paton matrices coincide with the generators of the gauge
group. A similar check can be performed in the $t$- and $u$-channels.
This confirms that we have correctly chosen the phases of the six
permutations that contribute to the amplitude.

We can now specialize to the case where the fermions are the
$U(1)$ gauginos by using the appropriate Chan-Paton matrices:
\begin{equation}
\lambda^{1} = \lambda^{2} = \frac{1}{\sqrt{N}}\ \id_{N}\, .
\label{chanpaton}
\end{equation}
The point-particle limit now vanishes and the interaction results
from purely ``stringy'' effects with massive string modes as
intermediate states. Inserting~(\ref{chanpaton}) in~(\ref{AALL})
and using again the expansion~(\ref{Bexp}), we obtain the first
correction to the QFT amplitude:
\begin{equation}
\mathcal{A}^{(2)}(\lambda_{L} \lambda_{R} \rightarrow B B) =
- \frac{2i \pi^2 {\alpha'}^{2} g^{2}_{YM}}{N} (2\pi)^{4}
\delta^{(4)}(\sum_{i}k_{i}) \Tr({\lambda^{3} \lambda^{4}})
K_{GB}(1,2,3,4)\, .
\label{AALLexp}
\end{equation}

\subsection{Interactions with scalars}

An analogous calculation can be done for the interaction between
the scalars and Goldstinos. The scalars are just the components of the
ten-dimensional gauge bosons in the transverse and internal
directions (in the latter case we need to take the zero-modes of the Kaluza-Klein expansion). The vertex operators can be obtained from~(\ref{GBVertex}) simply by choosing the gauge bosons to be
polarized in these directions. For example, we can obtain one such complex
scalar ($\phi^{(1)}$) by choosing the gauge boson to be polarized in the
45-direction:
\begin{equation}
\epsilon = \frac{1}{\sqrt{2}}(0,0,0,0,1,-i, 0,0,0,0)\, .
\end{equation}
Using this polarization in (\ref{GBVertex}) and bosonizing the fermions,
we obtain
\begin{eqnarray}
\nu^{(1)}_{-1}(x,k)  & = & \sqrt{2\alpha'} g_{YM}\ e^{-\phi} e^{-i
H_{1}} e^{ikX} \nonumber \\
\nu^{(1)}_{0}(x, k)  & = & 2g_{YM}\ (\ \frac{i}{\sqrt{2}} (\partial X^{4} -
i \partial X^{5}) + \alpha' (k\psi)e^{-i H_{1}} ) e^{ikX} \label{scalarVertex}
\end{eqnarray}
The CPT conjugate scalar has then polarization vector $\epsilon^{*}$.
Its vertex operator is obtained by reversing the internal helicity $e^{-i
H_{1}} \rightarrow e^{i H_{1}}$ and taking the complex conjugation
$iX^{5}\rightarrow -iX^{5}$. Similarly, the scalars $\phi^{(2)}$ and
$\phi^{(3)}$ correspond to gauge bosons polarized in the 67- and
89-planes, respectively, and their vertices $\nu^{(2)}_{-1}$ and
$\nu^{(3)}_{-1}$ are obtained just like~(\ref{scalarVertex}).

The calculation now proceeds in exactly the same way as for the gauge
bosons and the result is the same as in~(\ref{AALL}) except that we
need to replace the kinematic factor
\begin{equation}
K_{GB}(1,2,3,4) \rightarrow K_{S}(1,2,3,4) = u\ u^{T}_{L1} C \slash{k_{4}}
u_{R2}\, .
\label{SS_kinematic_factor}
\end{equation}
Inserting the Chan-Paton factors~(\ref{chanpaton}), we can expand in
powers of $\alpha'$, retaining only the first stringy correction. This
yields
\begin{eqnarray}
\mathcal{A}^{(2)}(\lambda_{L} \lambda_{R} \rightarrow \phi^{(i)}
\bar{\phi}^{(j)}) & = & -i (\frac{2 \pi^2 {\alpha'}^2 g^{2}_{YM}}{N}) (2\pi)^{4} \delta^{(4)}(\sum_{i}k_{i}) \nonumber  \\ & \times & \Tr({\lambda^{3} \lambda^{4}})\delta^{ij}
K_S(1,2,3,4)\, . \label{GGSS_untwisted}
\end{eqnarray}

\subsection{Interactions with fermions}

The interactions of two fermions with two Goldstinos of opposite helicity were studied in
ref.~\cite{Antoniadis:2001pt}. It was found that there are two
inequivalent cases, depending on the internal helicities of the fermions.
In the next section we shall recover these results by computing the
corresponding interaction on intersecting D-branes and taking the limit
of coincident branes. Here, we simply quote the result:
\begin{eqnarray}
\mathcal{A}^{(2)}(\lambda_{L} \lambda_{R} \rightarrow  f_{L} \bar{f}_{R}) & = &
-2i (\frac{2 \pi^2 {\alpha'}^2 g^{2}_{YM} }{N}) \nonumber \\
& \times & (2\pi)^{4} \delta^{(4)}(\sum_{i}k_{i}) \Tr({\lambda^{3}
\lambda^{4}}) K_F(1,2,3,4)\, ,
\label{GGFF_untwisted}
\end{eqnarray}
where the kinematic factor is
\begin{equation}
K_F(1,2,3,4) = \begin{cases} \frac{tu+su}{2s} & \text{case I} \\
\frac{tu}{2s} & \text{case II} \end{cases} \label{four_fermions1}
\end{equation}
Case I corresponds to the amplitude where the left-handed fermion $f_L$ has internal helicities $(-,-,-)$, whereas in case II it has
mixed internal helicities $(+,+,-),(+,-,+)$ or $(-,+,+)$.

In order to study the effective operator~(\ref{O4}) in string
theory, we also need to compute analogous four-fermion amplitudes involving
two Goldstinos with the same helicity. If we take the (incoming) Goldstinos to be
left-handed, then conservation of internal helicitiy requires the outgoing
fermions to be left-handed as well, with internal helicities $(-,-,-)$. This
corresponds to case I. The computation is straightforward. The vertex
operators for both the Goldstino and the fermions are given by~(\ref{GoldVertex}) together with appropriate Chan-Paton factors. The final result is:
\begin{equation}
\mathcal{A}^{(2)}(\lambda_L \lambda_L \rightarrow f_L f_L ) = -2 i (\frac{2 \pi^2
  {\alpha'}^2 g^2_{YM}}{N})(2\pi)^4 \delta^4(\sum_i k_i) \Tr(\lambda^3
\lambda^4) K'_F(1,2,3,4) \label{GGFF_untwisted_b}
\end{equation}
where
\begin{equation}
K'_F(1,2,3,4) = \begin{cases} \frac{s}{2} & \text{case I} \\
0 & \text{case II} \end{cases} \label{four_fermions1_b}
\end{equation}

\section{Intersecting D-branes} \label{intersecting_stacks}
\setcounter{equation}{0}
Intersecting D-branes have been used in recent years to construct
semi-realistic models with particle spectra and gauge groups close to the
Standard Model (see for example ref.~\cite{Uranga:2003pz} and references
therein). It is therefore interesting to examine the dynamics of
Goldstinos in this framework.

We consider the case of two stacks of D-branes intersecting in some
$d$-dimensional world-volume. The two stacks will be denoted by D$p$ and
D$p'$ with $N$, $N'$ being the respective number of branes in each
stack. The gauge group
is then $U(N)\times U(N')$ and the massless spectrum is divided into
several sectors
transforming in different representations of the gauge group:

\begin{itemize}
\item Open strings with both endpoints located on the same stack.
The corresponding excitations will (upon compactification) fill out
vector multiplets of ${\cal N}=4$ supersymmetry. Strings with both ends on D$p$
will transform in the adjoint representation of $U(N)$ and in the
trivial representation of $U(N')$ while those with both ends on D$p'$
will be in the trivial of $U(N)$ and in the adjoint of $U(N')$.

\item Open strings stretched between the two stacks. The corresponding states are localized at the intersection and transform in bifundamental representations of the gauge group. Specifically, an open string with its $\sigma=0$ end on the D$p$ and its $\sigma=\pi$ end on the D$p'$ ($pp'$ string) transforms in
$(N,\bar{N'})$. Here $\sigma$ is the world-sheet coordinate along the
open string. By interchanging the endpoints we obtain a $p'p$ string
transforming in $(\bar{N}, N')$. If the two stacks are localized at the
same position in the transverse directions, then the intersection will
carry massless fermions. For special values of the intersection angles
corresponding to unbroken supersymmetry (locally), the intersection will
also carry massless scalars which combine with the fermions into supermultiplets.
For generic angles, however, the scalars are massive or
tachyonic and the full massless spectrum is non-supersymmetric.
\end{itemize}

\subsection{Interactions with chiral fermions}
We will be interested in the interactions between Goldstinos and the
fermions located on the intersection. The study of non-linear
supersymmetry in Section~\ref{couplings_to_fermions} has shown that,
besides the coupling to the energy-momentum tensor, there are some
possible additional four-fermion interactions at order $\kappa^{2}$. We
wish to determine whether these coupling do appear in string theory for
generic D-brane intersection angles.

For simplicity, we shall specialize to the case of two stacks of D6-branes
intersecting in a (3+1)-dimensional world-volume. In such a configuration,
there can be no common transverse directions and the spectrum will generically
contain massless fermions.  As noted
in ref.~\cite{Berkooz:1996km}, the GSO projection requires these fermions
to be chiral. Furthermore, by choosing the intersection angles
appropriately, it is possible to preserve at most one supersymmetry\footnote{Of course,
by setting some intersection angle(s) to zero, it is possible to preserve more
supersymmetry. However in this case, from the four-dimensional point
of view, the fermions on the intersection will not be chiral anymore. We therefore take all intersection angles to
be non-vanishing.}. In
this case, there will be two (CPT conjugate) massless scalars on the
intersection (see subsection
\ref{section_twisted scalars}) which combine with
the fermions into chiral multiplets.

We start with two parallel stacks of D6-branes which we take to be
oriented in the 0123468 directions. We
then rotate the $D6'$ stack in the 45, 67 and 89 planes by angles $\phi_{1}$,
$\phi_{2}$ and $\phi_{3}$, respectively. The two stacks then intersect in the
0123 directions. It is convenient to introduce complex
coordinates:
\begin{equation}
Z^{1} = X^{4} + i X^{5} \qquad Z^{2} = X^{6} + i X^{7} \qquad Z^{3} =
X^{8} + i X^{9}
\end{equation}
Consider now a $66'$ string. In terms of the coordinates $Z^{k}$, it
will satisfy the following
boundary conditions:
\begin{eqnarray}
{\rm Re}(\partial_{\sigma} Z^{k}) = 0& \qquad {\rm Im}(Z^{k})= 0&
\qquad {\rm at}\qquad \sigma=0\nonumber \\ \\
{\rm Re}(e^{- i \phi_{k}}\partial_{\sigma} Z^{k}) = 0& \qquad {\rm
Im}(e^{-i \phi_{k}}Z^{k}) = 0& \qquad {\rm at}\qquad \sigma = \pi\nonumber
\end{eqnarray}
They imply that the mode expansions of $\partial Z^{k}$ and $\partial
\bar{Z}^k$ have mode numbers
shifted by $\frac{\phi_{k}}{\pi}$:
\begin{equation}
\partial Z^{k}(e^{ 2 i \pi} z) = e^{2i\phi_{k}} \partial Z^{k}(z)
\qquad \partial \bar{Z}^{k}(e^{2 i \pi} z) = e^{-2 i \phi_{k}}
\partial \bar{Z}^{k}(z) \label{periodicity1}
\end{equation}
where $z$ is the complex world-sheet coordinate.

To construct the vertex operators which create the $66'$ states we can
proceed in analogy with strings on orbifolds. Indeed, bosonic fields
satisfying the periodicity condition~(\ref{periodicity1}) can be thought
of as belonging to the twisted sector of an orbifold theory. Let us first
assume that the rotation angle $\phi_k$ is positive.
Following ref.~\cite{Dixon:1986qv}, the correct boundary condition near
the insertion of the vertex operator is implemented by means of a twist
field $\sigma_+$ which has an appropriate operator product expansion with
$\partial Z^{k}$ and $\partial \bar{Z}^{k}$:
\begin{equation}
\partial Z^{k}(z) \sigma_{+}^{k}(0) \sim z^{-(1- \frac{\phi_k}{\pi})}
\tau_{+}^{k}(0) \qquad
\partial \bar{Z}^{k}(z) \sigma_{+}^{k}(0) \sim z^{-
\frac{\phi_{k}}{\pi}} {\tau'}_{+}^{k}(0) \label{OPE1}
\end{equation}
where $\tau_{+}$ and $\tau'_{+}$ are ``excited'' twist fields. The conformal
dimension of $\sigma^{k}_{+}$ is $h_{\sigma} = \frac{1}{2}
\frac{\phi_{k}}{\pi}(1 - \frac{\phi_k}{\pi})$.

By superconformal symmetry, twisting the bosonic fields requires a
similar twisting of their world-sheet superpartners. Let $\psi^{k}$ and
$\bar{\psi}^{k}$ be the complexified fermions which are the superpartners
of $\partial Z^{k}$ and $\partial \bar{Z}^{k}$. World-sheet supersymmetry
and condition~(\ref{periodicity1}) require that, in the Ramond sector,
they satisfy
\begin{equation}
\psi^{k}(e^{2 i \pi} z) = - e^{ 2 i \phi_k} \psi^{k}(z) \qquad
\bar{\psi}^{k}(e^{ 2 i \pi} z) = - e^{- 2 i \phi_k} \bar{\psi}^{k}(z)
\label{periodicity2}
\end{equation}
which insures that the world-sheet supercurrent is anti-periodic on the plane (or periodic on the cylinder). We
therefore need to insert along with
$\sigma^{k}_{+}$ a fermionic twist field $S^{k}_{+}$ such that
\begin{equation}
\psi^{k}(z) S^{k}_{+}(0) \sim z^{\frac{\phi_k}{\pi} - \frac{1}{2}}
{t'}^{k}_{+}(0) \qquad
\bar{\psi}^{k}(z) S^{k}_{+}(0) \sim z^{\frac{1}{2} -
\frac{\phi_k}{\pi}} {t}^{k}_{+}(0) \label{OPE2}
\end{equation}
In terms of the bosonized fermions $H_{k}$, the fermionic twist fields
can be expressed as $S^{k}_{+} = e^{ i(\frac{\phi_k}{\pi} - \frac{1}{2})
H_k}$. Their conformal dimension is $h_{S} =
\frac{1}{2}(\frac{\phi_{k}}{\pi} -\frac{1}{2})^{2}$ and therefore
$h_{\sigma} + h_{S} = \frac{1}{8}$ independent of the rotation angle.
Moreover, ${t'}^k_+ = e^{ i(\frac{\phi_k}{\pi} + \frac{1}{2})H_k}$ and
${t}^k_+ = e^{ i(\frac{\phi_k}{\pi} - \frac{3}{2})H_k}$.

In the context of open strings on intersecting D-branes, the twist
fields can be thought of as implementing discrete changes of boundary
conditions. As one moves along the boundary of the world-sheet, crossing
a twist field amounts to moving (in the target space) across an
intersection from one stack to another.

The case of a negative rotation angle is similar. Instead of~(\ref{periodicity1})
and~(\ref{periodicity2}) we now have similar
expressions with $\phi_{k} \rightarrow - \phi_{k}$\footnote{From now on,
we shall always take $\phi_k$ to denote the magnitude of the rotation
angle and $\epsilon_k=\pm$ to denote its sign.}. This boundary condition
is implemented by means of anti-twist fields $\sigma^{k}_{-}$ and
$S^{k}_{-}$. The corresponding expressions for the operator product
expansions and conformal weights are obtained simply by making the
replacement
$\frac{\phi_k}{\pi} \rightarrow (1 - \frac{\phi_k}{\pi})$. In
particular, the conformal weights of twist and anti-twist fields are the
same.

The vertex operators for the massless fermions on the intersection can be
obtained from~(\ref{GoldVertex}) by replacing the bosonized spin
fields in the 45, 67 and 89 planes by the appropriate twist and anti-twist
fields. Consider first the case of a $66'$ string; we obtain:
\begin{equation}
\nu^{66'}_{-\frac{1}{2}} = g_{0} \lambda^{66'} e^{-\frac{\phi}{2}}
u_{\alpha} \Theta^{\alpha}
\prod_{k=1}^{3}(S^{k}_{\pm}\sigma^{k}_{\pm}) e^{ipX}\, . \nonumber
\end{equation}
The Chan-Paton factor $\lambda^{66'}$ is a $(N+N') \times (N+N')$ matrix
with entries only in the upper off-diagonal block in accordance with the
fact that a $66'$ string transforms in the $(N,\bar{N}')$ representation of $U(N) \times U(N')$. The sign of the
twist $\sigma^k_{\pm}S^k_{\pm}$ is given by the sign of the rotation in
the corresponding plane.  These signs determine via the GSO projection
the chirality of the Weyl spinor $u_{\alpha}$.

The $6'6$ string is obtained by interchanging the endpoints of the $66'$
string. This amounts to flipping the signs of all twists and replacing
the Chan-Paton matrix $\lambda^{66'}$  by one with non-vanishing entries
only in the lower off-diagonal block:
\begin{equation}
\bar{\nu}^{6'6}_{-\frac{1}{2}} = g_{0} \lambda^{6'6} e^{-\frac{\phi}{2}}
v_{\alpha} \Theta^{\alpha}
\prod_{k=1}^{3}(S^{k}_{\mp}\sigma^{k}_{\mp}) e^{ipX}\, . \nonumber
\end{equation}
This string transforms in the conjugate representation
$(\bar{N},{N'})$. In this sense,
the $6'6$ string can be viewed as the CPT conjugate of the $66'$
string. Furthermore, the GSO projection requires $v_{\alpha}$ to be a
Weyl spinor of chirality opposite to $u_{\alpha}$.

We now have to distinguish between two types of intersections. Let
$\epsilon_{1,2,3}$  denote the sign of the rotation in the 45, 67 and 89
plane, respectively. We shall call the product
$\epsilon_1 \times \epsilon_2 \times \epsilon_3$ the chirality of the
intersection. We then have:

\begin{enumerate}
\item Positive intersections \newline
For this type of intersection, the GSO projection requires
(conventionally) $u_{\alpha}$ to be left-handed and $v_{\alpha}$ to be
right-handed. This implies that, in our convention, positive intersections carry only negative helicity particles ($66'$ strings) and positive
helicity anti-particles ($6'6$ strings).
\item Negative intersections \newline
This is precisely the opposite situation. The GSO projection requires
$u_{\alpha}$ and $v_{\alpha}$ to be right- and left-handed,
respectively. Therefore negative intersections carry only positive
helicity particles and negative helicity anti-particles\footnote{Positive (negative) chirality intersections make positive
(negative) contributions to the intersection number $I_{66'}$ as
defined in ref.~\cite{Uranga:2003pz}.}.
\end{enumerate}

For our purposes it will be sufficient to specialize to the case of
positive intersections. The vertex operators are then given by
\begin{eqnarray}
\nu^{66'}_{-\frac{1}{2}} & = & g_{0} \lambda^{66'}
e^{-\frac{\phi}{2}} u_{L\alpha} \Theta^{\alpha}
\prod_{k=1}^{3}(S^{k}_{\epsilon_k}\sigma^{k}_{\epsilon_k}) e^{ipX} \nonumber \\
\bar{\nu}^{6'6}_{-\frac{1}{2}} & = & g_{0} \lambda^{6'6}
e^{-\frac{\phi}{2}} u_{R\alpha} \Theta^{\alpha}
\prod_{k=1}^{3}(S^{k}_{-\epsilon_k}\sigma^{k}_{-\epsilon_k}) e^{ipX}
\label{FermionVertex}
\end{eqnarray}
where $\epsilon_1 \epsilon_2 \epsilon_3 = +1$ and
\begin{equation}
\lambda^{66'} = \begin{pmatrix} 0 & t \\ 0 & 0 \end{pmatrix} \qquad
\lambda^{6'6} = \begin{pmatrix} 0 & 0 \\ t' & 0 \end{pmatrix}\, .
\label{chan_paton_fermions}
\end{equation}
The constant $g_0$ is a normalization which can be expressed in terms of the
Yang-Mills couplings $g_{YM}$ and $g_{YM}'$ by comparing the
point-particle limit of string amplitudes with the corresponding field
theory calculation. We can for example compute the three-point function of
two fermions and a gauge boson of $U(N)$. This is given by
\begin{equation}
\mathcal{A}(1,2,3) = A(1,2,3)\ \Tr(\lambda^{1} \lambda^{2} \lambda^{3}) + (2
\leftrightarrow 3)
\end{equation}
with
\begin{equation}
A(1,2,3) = iC_{D}<c\nu^{66'}_{-\frac{1}{2}}(0,p_1, u_{L1})
c\bar{\nu}^{6'6}_{-\frac{1}{2}}(1, p_2, u_{R2}) c\nu_{-1}(\infty, p_3,
\epsilon)>\, .
\end{equation}

Inserting the vertex operators~(\ref{FermionVertex}) and~(\ref{GBVertex}) and
using~(\ref{DiscConstant}) for the constant $C_{D}$, we obtain
\begin{equation}
\mathcal{A}(1,2,3) = i \frac{g_{0}^2}{2 {\alpha'}^{\frac{3}{2}} g_{YM} V_{c}}
(u_{1L}^{T}C \slash{\epsilon}\ u_{2R} )(2\pi)^{4} \delta^{4}(\sum_{i}p_i)
\Tr([\lambda^{1}, \lambda^{2}] \lambda^{3})\, .
\label{three_point}
\end{equation}
Here we have used the correlator of twist fields which is completely
determined (up to a normalization which we fix to unity) by conformal
invariance and the conformal weights $h_{\sigma}$, $h_{S}$:
\begin{equation}
<(S^{k} \sigma^{k})_{+}(x_1) (S^{k} \sigma^{k})_{-}(x_2)> =
x_{12}^{-\frac{1}{4}}\, .
\end{equation}
Also, the correlator of exponentials is given by
\begin{equation}
<\prod_i e^{ip_i X}> = (2\pi)^4 \delta^4(\sum_i p_i) \prod_{i<j}
x_{ij}^{2\alpha' p_i p_j}\, .
\end{equation}
Note that the factor $V_c$ appearing in (\ref{correlator_exp}) is now
missing. This is because the fields $Z^{k}$ and $\bar{Z}^{k}$ do not have
any zero modes over which we have to integrate since the strings are localized
on the intersection. It follows that there is
an extra factor of $V^{\frac{1}{2}}_{c}$ in the normalization constant
$g_0$. Indeed, comparing (\ref{three_point}) with the corresponding
vertex in Yang-Mills theory we obtain
\begin{equation}
g_{0} = (4 {{\alpha'}^3})^{\frac{1}{4}} g_{YM} V_{c}^{\frac{1}{2}}\, .
\end{equation}
If instead we had chosen a gauge boson of $U(N')$ we would have
obtained the same result with $g_{YM} \rightarrow g'_{YM}$ and $V_{c}
\rightarrow V'_{c}$. In particular, this
implies the relation
\begin{equation}
\frac{g_{YM}}{g_{YM}'} = (\frac{V'_c}{V_c})^{\frac{1}{2}}\, ,
\label{gym_relation}
\end{equation}
which follows from the fact that both Yang-Mills couplings are
related to the string coupling $g_{s}$ by dimensional reduction.

We need again to identify the Goldstinos in the spectrum. In contrast to
the case of a single stack of D-branes, there is now an additional
complication from the fact that the amount of broken supersymmetry, and therefore the
number of Goldstinos, depends on the intersection angles. As mentioned in
the previous section, the Goldstinos should be gauge singlets and
therefore must be identified with the gauginos of the
$U(1)$ factors in $U(N)$ and $U(N')$. This would result in a total of 16
Goldstinos which is in agreement with Goldstone's theorem in the case of
non-supersymmetric configurations. However, if some unbroken
supersymmetry remains, then not all U(1) gauginos can be identified with
Goldstinos. Instead, it turns out that in this case the physical
Goldstinos correspond to certain linear combinations of $U(1)$ gauginos.
The coefficients appearing in these linear combinations can be determined by the
condition that the Goldstinos, being gauge neutral, do not interact with the chiral fermions in the
QFT limit $\alpha' \rightarrow 0$.

The vertex operator for the D$6$ gaugino is given in (\ref{GoldVertex}). It needs
to be supplemented by the appropriate Chan-Paton factor $\lambda$ which in the present
context is an
$(N+N')\times(N+N')$ matrix with the identity in the upper diagonal
block and all other entries vanishing:
\begin{equation}
\lambda = \frac{1}{\sqrt{N}} \begin{pmatrix} \id_{N} & 0 \\ 0 & 0
\end{pmatrix}\, . \label{chan_paton_goldstino1}
\end{equation}

To obtain the vertices for the D$6'$ gauginos, we substitute
$g_{YM} \rightarrow g'_{YM}$ and $\lambda \rightarrow \lambda'$ where
$\lambda'$  has the identity in the lower diagonal block. Also, we have
to replace $H_{k}$ by $H'_k = H_{k} + \epsilon_k \phi_k$. This shift
comes from the rotation of the D$6'$ stack relative to the D$6$. Indeed,
$e^{\pm iH_{k}}$ is the bosonization of $\frac{\psi^{2k+2} \pm
i\psi^{2k+3}}{\sqrt{2}}$ while  $e^{\pm iH'_{k}}$ is the bosonization of
$\frac{{\psi'}^{2k+2} \pm i{\psi'}^{2k+3}}{\sqrt{2}}$. The latter is
obtained from the former by applying a rotation by an angle $\phi_k$ in a
positive ($\epsilon_k = +1 $) or negative ($\epsilon_k = -1$) direction:
\begin{equation}
\frac{{\psi'}^{2k+2} \pm i{\psi'}^{2k+3}}{\sqrt{2}} =
e^{\pm i \epsilon_k \phi_k} \frac{{\psi}^{2k+2} \pm
i{\psi}^{2k+3}}{\sqrt{2}}\, .
\end{equation}
In terms of the fields $H_k$, this rotation corresponds to a shift by
$\epsilon_k \phi_k$ . Putting everything together, the vertex operators
for the D$6'$ gauginos become
\begin{eqnarray}
{\nu'}^{G}_{-\frac{1}{2}}(x, k, u_{L}) &=& (4
{\alpha'}^{3})^{\frac{1}{4}} g'_{YM} e^{-i\frac{\Phi}{2}} \lambda' \
u_{L\alpha}  \Theta^{\alpha} \
e^{-\frac{\phi}{2}} e^{ -\frac{i}{2}(H_{1} +
   H_{2} + H_{3})}\ e^{ipX} \nonumber  \\
\bar{\nu'}^{G}_{-\frac{1}{2}}(x, k, u_{R}) &=&  (4
{\alpha'}^{3})^{\frac{1}{4}} g'_{YM} e^{i\frac{\Phi}{2}} \lambda' \
u_{R\alpha} \Theta^{\alpha}\
e^{-\frac{\phi}{2}}e^{ \frac{i}{2}(H_{1} +
   H_{2} + H_{3})}\ e^{ipX} \label{GauginoVertex}
\end{eqnarray}
where we have defined $\Phi$ as the sum of rotation angles
\begin{equation}
\Phi = \epsilon_1 \phi_1 + \epsilon_2 \phi_2 + \epsilon_3 \phi_3
\label{sum_of_angles}
\end{equation}
and
\begin{equation}
\lambda' = \frac{1}{\sqrt{N'}} \begin{pmatrix} 0 & 0 \\ 0 & \id_{N'}
\end{pmatrix}\, . \label{chan_paton_goldstino2}
\end{equation}
In~(\ref{sum_of_angles}) we take into account the respective signs of
the rotations.

It turns out that the physical Goldstino with internal helicities ${- - -}$ and its CPT conjugate are given
by the following linear combinations of gauginos:
\begin{eqnarray}
\nu^{G_{phys}}_{-\frac{1}{2}} & = & \frac{1}{\sqrt{N V_c + N' V'_c}}
(\sqrt{N V_c}\ \nu_{-\frac{1}{2}}^{G} + \sqrt{N' V'_c}\
{\nu'}^{G}_{-\frac{1}{2}}) \nonumber \\
\bar{\nu}^{G_{phys}}_{-\frac{1}{2}} & = & \frac{1}{\sqrt{N V_c + N' V'_c}}
(\sqrt{N V_c}\ {\bar{\nu}}_{-\frac{1}{2}}^{G} + \sqrt{N' V'_c}\
\bar{\nu'}^{G}_{-\frac{1}{2}}) \label{physical_goldstino}
\end{eqnarray}
As we shall see, in the QFT limit, the physical Goldstino does not
interact with  the massless ``matter'' and gauge fields. In addition, in
the limit $\phi_k \rightarrow 0$ (\ref{physical_goldstino}) reduces to
(\ref{GoldVertex}).

The scattering amplitude $\mathcal{A}(\lambda_{L} \lambda_{R}
\rightarrow f_L \bar{f}_{R})$  is obtained from (\ref{totamplitude}) and
(\ref{amplitude}) and has four contributions:
\begin{equation}
\mathcal{A} = \frac{ NV_c \mathcal{A}^{66} + N' V'_c\mathcal{A}^{6'6'}
+ \sqrt{NN' V_cV'_c} \mathcal{A}^{66'} + \sqrt{NN' V_c V'_c}
\mathcal{A}^{6'6}}{NV_c + N'V'_c}\, . \label{gaugino_sum}
\end{equation}
The superscripts of the amplitudes on the right-hand side indicate which
gauginos are involved. For instance, according to~(\ref{totamplitude})
and~(\ref{amplitude}),  to evaluate $\mathcal{A}^{66}$ we need the
correlation function on the disc:
\begin{equation}
<\nu^{66}_{-\frac{1}{2}} \bar{\nu}^{66}_{-\frac{1}{2}}
\bar{\nu}^{6'6}_{-\frac{1}{2}} \nu^{66'}_{-\frac{1}{2}} >\, .
\end{equation}
Inserting the vertex operators, this factorizes into a product of simple
correlators and we obtain
\begin{equation}
\begin{split}
<\nu^{66}_{-\frac{1}{2}} \bar{\nu}^{66}_{-\frac{1}{2}}
\bar{\nu}^{6'6}_{-\frac{1}{2}}
\nu^{66'}_{-\frac{1}{2}}> & =  (4 {\alpha'}^{3}) g^{4}_{YM} V_{c} (2\pi)^4
\delta^4(\sum_i p_i) (u^T_{1L}Cu_{4L})(u^T_{2R}Cu_{3R})\\
&\times \prod_{i<j}x_{ij}^{2\alpha'p_i p_j}
\{ x^{-1}_{12} x^{\frac{\Phi}{2\pi} - \frac{\epsilon-1}{2}}_{13}
x^{-\frac{\Phi}{2\pi}
+ \frac{\epsilon+1}{2}}_{14}  x^{-\frac{\Phi}{2\pi} +
\frac{\epsilon-1}{2}}_{23}
x^{\frac{\Phi}{2\pi} - \frac{\epsilon+1}{2}}_{24} \}\, ,
\end{split}
\end{equation}
where we have defined the quantity
\begin{equation}
\epsilon = \frac{\epsilon_1 +
\epsilon_2 +  \epsilon_3 - 1}{2}\, ,
\end{equation}
which equals either $+1$ or $-1$ for positive intersections. Using this correlator in~(\ref{amplitude}), one finds the function $A(1,2,3,4)$. Including also the other two permutations of vertex operators according
to~(\ref{totamplitude}), we end up with
\begin{equation}
\begin{split}
\mathcal{A}^{66} &(1,2,3,4) = 2i \alpha' g^2_{YM} (2\pi)^{4}
\delta^4(\sum_i p_i)
(u^T_{1L}Cu_{4L})(u^T_{2R}Cu_{3R}) \\ & \biggl(\ \frac{\Gamma(-\alpha' s)
\Gamma(-\alpha'u - \frac{\Phi}{2\pi} +
\frac{\epsilon+1}{2})}{\Gamma(-\alpha's - \alpha'u - \frac{\Phi}{2\pi}
+ \frac{\epsilon+1}{2})}\
\Tr(\lambda^{1}\lambda^{2}\lambda^{3}\lambda^{4} +
\lambda^{4}\lambda^{3}\lambda^{2}\lambda^{1})
\\ &  \pm \frac{\Gamma(-\alpha' t + \frac{\Phi}{2\pi} -
\frac{\epsilon-1}{2}) \Gamma(-\alpha'u - \frac{\Phi}{2\pi} +
\frac{\epsilon+1}{2})}{\Gamma(-\alpha't - \alpha'u +1)}
\  \Tr(\lambda^{1}\lambda^{3}\lambda^{2}\lambda^{4} +
\lambda^{4}\lambda^{2}\lambda^{3}\lambda^{1})
\\ &  - \frac{\Gamma(-\alpha' s) \Gamma(-\alpha't + \frac{\Phi}{2\pi}
- \frac{\epsilon-1}{2})}{\Gamma(-\alpha's - \alpha't +
\frac{\Phi}{2\pi} - \frac{\epsilon-1}{2})}
\  \Tr(\lambda^{1}\lambda^{2}\lambda^{4}\lambda^{3} +
\lambda^{3}\lambda^{4}\lambda^{2}\lambda^{1})\ \biggr)\, .
\label{goldstino_amplitude}
\end{split}
\end{equation}

Note that, generically, there are no massless poles in either the $t$- or
$u$-channels. This is expected since the intermediate states in these
channels are twisted and correspond  to scalars ($u$-channel) or vectors
($t$-channel) on the intersection, both of which are massive or tachyonic
for generic values of the intersection angles. On the other hand, the
massless intermediate state in the $s$-channel  is a gauge boson.
Therefore, there must be a pole except if the external fermions are
neutral. In this case, the $s$-channel poles coming from the first and
third term in~(\ref{goldstino_amplitude}) must cancel and this fixes the
negative sign of the third term. To fix the sign of the second term we go
to the limit $\phi_{k} \rightarrow 0$ where the two stacks become
coincident. If $\epsilon=1$, there is a massless pole in the $t$-channel
corresponding to an intermediate gauge boson. Since the fermions are then
in the adjoint representation, this pole must be proportional to
$\Tr([\lambda^1, \lambda^3][\lambda^2,
\lambda^4])$, fixing the sign of the second term to be positive. By a
similar reasoning, it can be easily seen that for the case $\epsilon =
-1$, the sign of the second term must be negative.

The appropriate Chan-Paton matrices are $\lambda^{1} = \lambda^{2} =
\lambda$, $\lambda^{3} = \lambda^{6'6}$ and $\lambda^{4} =
\lambda^{66'}$. Using the explicit expressions~(\ref{chan_paton_fermions})
and~(\ref{chan_paton_goldstino1}), we see
that  only the first and third term contribute. The second term
corresponds to orderings of the vertex operators where the chiral
fermions lie between the gaugino vertices on the world-sheet boundary,
implying that the gauginos must be located on different stacks.

The other gaugino amplitudes can be obtained directly from~(\ref{goldstino_amplitude}).
 Thus, to obtain
$(\mathcal{A}^{6'6'},\ \mathcal{A}^{66'},\ \mathcal{A}^{6'6})$ we have to
substitute
\begin{equation}
g^{2}_{YM} \rightarrow ({g'}^{2}_{YM},\ e^{i\frac{\Phi}{2}}g_{YM}g'_{YM},\
e^{-i\frac{\Phi}{2}}g_{YM}g'_{YM})
\nonumber
\end{equation}
and use the appropriate Chan-Paton matrices for the gauginos:
\begin{equation}
(\lambda^1, \lambda^2) = (\lambda', \lambda'), (\lambda, \lambda'),
(\lambda', \lambda),
\end{equation}
respectively. The phase factors in $\mathcal{A}^{66'}$ and
$\mathcal{A}^{6'6}$ are crucial in order to cancel any dimension 6
operators in the low-energy effective action.

Putting everything together in~(\ref{gaugino_sum}) and using the
relation~(\ref{gym_relation}), we finally obtain the full
Goldstino-fermion scattering amplitude:
\begin{equation}
\begin{split}
\mathcal{A}(\lambda_L \lambda_R \rightarrow f_{L} \bar{f}_R)  =
& - 2 i (\frac{2 \pi^2 {\alpha'}^2 g_{YM} g'_{YM} V^{\frac{1}{2}}_c
{V'}^{\frac{1}{2}}_c}{NV_c + N' V'_c}) \\ & (2\pi)^4 \delta^4(\sum_i p_i)
\Tr(tt') K_F(1,2,3,4)\, , \label{goldstino_totamplitude}
\end{split}
\end{equation}
where $K_F(1,2,3,4)$ is the kinematic factor
\begin{equation}
K_F(1,2,3,4) = - \frac{1}{\alpha'
\pi^2}(u^T_{1L}Cu_{4L})(u^T_{2R}Cu_{3R}) f_{\epsilon}(s,t,u)
\end{equation}
with
\begin{equation}
\begin{split}
f_{\epsilon}(s,t,u)  = & \frac{\Gamma(-\alpha' s) \Gamma(-\alpha'u -
\frac{\Phi}{2\pi} + \frac{\epsilon+1}{2})}{\Gamma(-\alpha's -
\alpha'u - \frac{\Phi}{2\pi} + \frac{\epsilon+1}{2})} -
\frac{\Gamma(-\alpha' s) \Gamma(-\alpha't + \frac{\Phi}{2\pi}
- \frac{\epsilon-1}{2})} {\Gamma(-\alpha's - \alpha't +
\frac{\Phi}{2\pi} - \frac{\epsilon-1}{2})} \\  +\epsilon &
cos(\frac{\Phi}{2})    \frac{\Gamma(-\alpha' t + \frac{\Phi}{2\pi} -
\frac{\epsilon-1}{2}) \Gamma(-\alpha'u - \frac{\Phi}{2\pi} +
\frac{\epsilon+1}{2})}{\Gamma(-\alpha't - \alpha'u + 1)}\, .
\label{function}
\end{split}
\end{equation}

Below we give the leading term in the expansion of this function for the
two cases $\epsilon = +1$ and $\epsilon = -1$.

\begin{enumerate}
\item[I.] All three rotations are positive ($\epsilon = 1$):
\begin{equation}
f_1(s,t,u) = \begin{cases} \alpha'u \frac{\pi^2}{2} & \text{if $\Phi=0$}, \\
  -\alpha' t \pi^2  - \alpha' s \frac{\pi^2}{2} & \text{if $\alpha's,
\alpha't, \alpha'u << |\Phi|$} \end{cases} \label{expansion1}
\end{equation}

\item[II.] One rotation is positive and two are negative ($\epsilon = -1$):
\begin{equation}
f_{-1}(s,t,u) = \begin{cases} - \alpha' t\frac{\pi^2}{2} & \text{if
$\Phi=0$}, \\
  -\alpha't \pi^2 - \alpha' s \frac{\pi^2}{2} & \text{if $\alpha's,
\alpha't, \alpha'u << |\Phi|$} \end{cases} \label{expansion2}
\end{equation}

\end{enumerate}
Combining the above expansions with the explicit expressions for the Weyl
spinors we obtain for the kinematic factor:

\begin{eqnarray}
{\rm I.} \qquad K_F(1,2,3,4) = \begin{cases} \frac{su + tu}{2s} &
\text{if $\Phi=0$}
\\ \frac{su + 2tu}{2s} & \text{if $\Phi \neq 0$} \end{cases}
\label{KFI} \\
{\rm II.} \qquad K_F(1,2,3,4) = \begin{cases} \frac{tu}{2s} & \text{if
$\Phi=0$} \\
\frac{su + 2tu}{2s} & \text{if $\Phi \neq 0$} \end{cases} \label{KFII}
\end{eqnarray}

As promised, the quantum field theory limit $\alpha' \rightarrow 0$ vanishes
and the first stringy correction corresponds to an effective dimension 8
operator. Furthermore, we see that the case $\Phi=0$ reproduces the
results of ref.~\cite{Antoniadis:2001pt} for a single stack of D-branes
that were given in~(\ref{four_fermions1}).~\footnote{Notice however the
difference of our result from the result of
ref.~\cite{Antoniadis:2001pt} in the ${\cal N}=2$ supersymmetric case of
orthogonal branes, where one rotation angle vanishes and the other two
equal $\pm\pi/2$.} Starting with the vertex operators~(\ref{FermionVertex})
and taking the limit of coincident stacks, we see
that in case I the fermions have internal helicities
$(+,+,+)$ and
$(-,-,-)$ while in case II they have mixed internal helicities.

\subsection{Interaction with twisted scalars} \label{section_twisted scalars}

As stated previously, generically the intersection does not carry
massless scalars. A notable exception is of course the case of
supersymmetric configurations. In the Neveu-Schwarz (NS) sector, the
zero-point energy for twisted states is
\begin{equation}
E_0 =  \sum_k \frac{\phi_k}{2\pi} - \frac{1}{2} \quad ;\quad 0 \leq \phi_k
\leq\frac{\pi}{2}\, .
\end{equation}
The lowest state, which we denote as $|0;p>$, has squared-mass $m^2 =
E_0$ and is typically tachyonic. However, it is expected to be projected
out by an appropriate GSO projection. The lowest lying scalars in the
physical spectrum are then:
\begin{equation}
|j> \equiv \bar{\psi}_{-\frac{1}{2} + \frac{\phi_j}{\pi}} |0; p>
\qquad j = 1,2,3 \label{massless_scalars}
\end{equation}
where $\bar{\psi}_{-r+\frac{\phi_j}{\pi}}$ are the creation operators
appearing in the mode expansion of $\bar{\psi}^j$ in the NS sector. Here
we have assumed that all rotations defining the intersection are
positive, i.e. $\epsilon_1= \epsilon_2 = \epsilon_3 = 1$. If some
$\epsilon_k=-1$, then we need to replace
$\bar{\psi}_{-r+\frac{\phi_j}{\pi}} \rightarrow
\psi_{-r+\frac{\phi_j}{\pi}}$.

The masses of the states (\ref{massless_scalars}) are
\begin{eqnarray}
m_1^2 & = & \frac{1}{2\pi \alpha'}(-\phi_1 + \phi_2 + \phi_3) \nonumber \\
m_2^2 & = & \frac{1}{2\pi \alpha'}(\phi_1 - \phi_2 +  \phi_3) \nonumber \\
m_3^2 & = & \frac{1}{2\pi \alpha'}(\phi_1 + \phi_2 - \phi_3) 
\label{mass_shell_conditions}
\end{eqnarray}
If all angles are non-vanishing, then at most one of these scalars can be
made massless and this happens precisely when one supersymmetry is restored. One
can then consider interactions of this scalar with the Goldstinos
in the low-energy effective action.

By conservation of the internal helicities, it is clear that the
operator~(\ref{O8}) cannot appear in the effective action. The remaining
operators involving scalars are~(\ref{O2}) and~(\ref{O6}), both of which
will contribute to the amplitude $\mathcal{A}(\lambda_L \lambda_R
\rightarrow \phi \bar{\phi})$. Since the computation of this amplitude in
string theory is independent of which of the scalars is massless, we
consider for definiteness a configuration where the massless one is
$|j=3>$.

To construct the vertex operator for this state, we start from the
tachyon vertex and act with $\bar{\psi}^3_{-\frac{1}{2} +
\frac{\phi_3}{\pi}}$. The construction of the tachyon vertex follows
closely the method discussed in the previous subsection for the fermion
vertices, except of course that the tachyon is in the NS sector. To
implement the boundary condition (\ref{periodicity1}) we insert a
(bosonic) twist field $\sigma^{k}_{+}$. By world-sheet supersymmetry,
this field must be accompanied by a fermionic twist field $S^k_{+} = e^{i
\frac{\phi_k}{\pi}H_k}$. The corresponding operator product expansion
(OPE) with the fields $\psi^k,\bar{\psi^k}$ is
\begin{equation}
\psi^{k}(z) S^{k}_{+}(0) \sim z^{\frac{\phi_k}{\pi}} {t'}^k_+(0) \qquad
\bar{\psi}^{k}(z) S^{k}_{+}(0) \sim z^{- \frac{\phi_k}{\pi}} t^k_+(0)
\label{OPE2_NS}
\end{equation}
where
\begin{eqnarray}
t^k_+ & = & e^{i(\frac{\phi_k}{\pi}-1)H_k} \nonumber \\
{t'}^k_+ & = & e^{i(\frac{\phi_k}{\pi}+1)H_k} \nonumber
\end{eqnarray}
This makes the supersymmetry current periodic on the plane (or anti-periodic on the cylinder), as is appropriate for the
NS sector. The full vertex operator for the $66'$ tachyon is then
\begin{equation}
\nu^0_{-1} = g_0 e^{-\phi} \lambda^{66'} \prod^{3}_{k=1}
(\sigma^k_{+} S^k_{+})\ e^{ipX}\, . \label{tachyon_vertex}
\end{equation}
where $g_0$ is some normalization constant.

The vertex for the massless scalar in the $(-1)$-ghost picture is then
given by
\begin{equation}
\nu^{(3)}_{-1}(0) = \bar{\psi}^{3}_{-\frac{1}{2} + \frac{\phi_3}{\pi}} \cdot
\nu^{0}_{-1}(0) = \oint dz\ z^{\frac{\phi_3}{\pi}-1} \bar{\psi}^3(z)
\nu^{0}_{-1}(0)\, .
\end{equation}
Using the OPEs~(\ref{OPE2_NS}), we see that the appropriate vertex
operator is obtained from (\ref{tachyon_vertex}) by replacing $S_+^3 \rightarrow t^3_+$:
\begin{equation}
\nu^{(3)}_{-1}(z)  =  g_0 e^{-\phi} \lambda^{66'} (\sigma^{1}_{+}
S^1_{+})(\sigma^{2}_{+} S^2_{+})(\sigma^{3}_{+} t^3_{+}) e^{ipX}\, .
\end{equation}
Conformal invariance requires this operator to have conformal weight
1, from which we recover the mass-shell condition~(\ref{mass_shell_conditions}).

In order to compute the 4-point amplitude, we also need the scalar
vertex operator in the $0$-picture. This is obtained by operating with
the supercharge $G_{-\frac{1}{2}}$ on $\nu^{3}_{-1}$ after removing the
superconformal ghost $e^{-\phi}$~\cite{Friedan:1985ge}:
\begin{equation}
\nu^{(3)}_0 (0) = Q_{BRST} \cdot \nu^{(3)}_{-1}(0) = \int dz e^{\phi} T_{F}(z) \nu^{(3)}_{-1}(0)\, . \label{picture_changing}
\end{equation}
{}From the OPEs~(\ref{OPE1}) and~(\ref{OPE2_NS}) we get
\begin{equation}
\begin{gathered}
G_{-\frac{1}{2}} \cdot (\sigma^k_{+}S^k_{+}) = i
\sqrt{\frac{2}{\alpha'}} \tau^k_{+} t^k_{+} \\
G_{-\frac{1}{2}} \cdot (\sigma^k_{+}t^k_{+})= i
\sqrt{\frac{2}{\alpha'}} {\tau'}^k_{+}S^k_{+} \\
G_{-\frac{1}{2}} \cdot e^{ipX} = \sqrt{\frac{2}{\alpha'}}
\alpha' (p \cdot \psi) e^{ipX}\, .
\end{gathered}
\end{equation}
which yields
\begin{eqnarray}
\nu^{(3)}_{0} & = & \sqrt{\frac{2}{\alpha'}} g_0 \lambda^{66'}\biggl( i( \tau^{1}_{+}
t^{1}_{+}\ \sigma^{2}_{+} S^{2}_{+}\ \sigma^{3}_{+} t^{3}_{+} +
\sigma^{1}_{+} S^{1}_{+}\ \tau^{2}_{+} t^{2}_{+}\ \sigma^{3}_{+}
t^{3}_{+} \nonumber \\ & + & \sigma^{1}_{+} S^{1}_{+}\ \sigma^{2}_{+} S^{2}_{+}\
{\tau'}^{3}_{+} S^3_{+} )
 + \prod^{2}_{k=1}(\sigma^{k}_{+} S^{k}_{+}) \sigma^3_+ t^3_+\alpha'
(p \cdot \psi)
   \biggr) e^{ipX}\, .
\end{eqnarray}
The first three terms do not contribute to the amplitude, since they
result in correlators of the form
\begin{equation}
<\sigma_+(z) \tau_+(z')> \qquad <\sigma_+(z) {\tau'_+}(z')>
\end{equation}
both of which vanish by conformal invariance. Effectively, the scalar
vertex then becomes
\begin{equation}
\nu^{(3)}_{0} \rightarrow \sqrt{\frac{2}{\alpha'}} g_0 \lambda^{66'}
\prod^{2}_{k=1}(\sigma^{k}_{+} S^{k}_{+}) \sigma^3_+ t^3_+ \alpha' (p
\cdot \psi) e^{ipX}\, .
\end{equation}

This construction easily generalizes to arbitrary twists with the result:
\begin{eqnarray}
\nu^{(3)}_{-1}(z) & = & g_0 e^{-\phi} \lambda^{66'}
\prod_{k=1}^{2}(\sigma^{k}_{\epsilon_k}
S^k_{\epsilon_k})(\sigma^3_{\epsilon_3} t^3_{\epsilon_3}) e^{ipX} \label{scalar_intersection} \\
\nu^{(3)}_{0} & \rightarrow & \sqrt{\frac{2}{\alpha'}} g_0
\lambda^{66'} \prod^{2}_{k=1}(\sigma^{k}_{\epsilon_k}
S^k_{\epsilon_k})\sigma^{3}_{\epsilon_3}t^{3}_{\epsilon_3} \alpha' (p
\cdot \psi) e^{ipX}
\end{eqnarray}
The vertex for the CPT conjugate state is obtained as usual by
flipping the signs of all twists and replacing the Chan-Paton matrix
$\lambda^{66'} \rightarrow \lambda^{6'6}$. Also, the normalization constant
$g_0$ can be deduced by the same method as for the chiral fermions with
the result $g_0 = \sqrt{2 \alpha'} g_{YM} V^{\frac{1}{2}}_c$. The calculation then
proceeds in exactly the same way as for the fermions. The final
result is
\begin{eqnarray}
\mathcal{A}(\lambda_L \lambda_R \rightarrow \phi^3 \bar{\phi}^3) & = & -
i (\frac{2 \pi^2 {\alpha'}^2 g_{YM}g'_{YM}V^{\frac{1}{2}}_c
{V'}^{\frac{1}{2}}_{c}}{N V_c + N' V'_c} ) \nonumber \\ & \times & (2\pi)^{4} \delta^4(\sum_k
p_i) \Tr(tt') K_S(1,2,3,4) \label{goldstino_scalar_amplitude}
\end{eqnarray}
where the kinematic factor $K_S(1,2,3,4)$ is given by
\begin{equation}
K_S(1,2,3,4) = \frac{2}{\alpha' \pi^2} u^T_{L1}C\slash{p_4}u_{R2}\
f_{\epsilon_3}(s,t,u)\, .
\end{equation}

Using the expansions for the function $f_{\epsilon}(s,t,u)$ given in~(\ref{expansion1}) and~(\ref{expansion2}), we obtain the first non-vanishing term in the expansion of $K_S$:

\begin{enumerate}
\item[I.] $\epsilon_3=1$ and the kinematic factor becomes
\begin{equation}
K_S(1,2,3,4) = u^T_{1L}C \slash{p_4} u_{2R}\ (u - t) \qquad (\Phi \neq
0) \label{KSI}
\end{equation}

\item[II.] Using~(\ref{sum_of_angles}) and the fact that the scalar is massless (i.e. $\phi_3 = \phi_1 + \phi_2$), it is easy to see that
$\Phi=0$ requires $\epsilon_3=+1$ while $\Phi \neq 0$ requires
$\epsilon_3 = -1$. We then obtain:
\begin{equation}
K_S(1,2,3,4) = u^T_{1L}C \slash{p_4} u_{2R} \begin{cases} u & \text{if
$\Phi=0$} \\ u-t & \text{if $\Phi\neq0$} \end{cases} \label{KSII}
\end{equation}
\end{enumerate}
The same result applies for any other of the scalars in
(\ref{massless_scalars}), as long as the angles are chosen to make
this particular scalar massless.

\subsection{Interactions with strings located on single D-brane stacks}

In Section \ref{single_stack}, we computed interaction amplitudes
involving gauginos, gauge bosons and adjoint fermions for the case of
a single stack of $N$ D-branes. The Goldstino was identified as the gaugino
of the $U(1)$ supermultiplett. These results can be immediately translated
into corresponding results for the case of two intersecting stacks of D-branes.
Indeed, in~(\ref{physical_goldstino}) we identified the Goldstino as a linear
combination of the $U(1)$ gauginos of the two stacks.

Consider for example interactions of the Goldstino with gauge
bosons of the D6 stack. The amplitude $\mathcal{A}(\lambda_L \lambda_R \rightarrow BB)$
has four contributions just like
the four-fermion amplitude~(\ref{gaugino_sum}) computed in the
previous subsection. However, due to the vanishing of the corresponding
traces of Chan-Paton matrices, only one of these contributions,
$\mathcal{A}^{66}$, survives. As expected, the $U(N)$ gauge bosons interact
only with the $U(1)$ gauginos of the $D6$ stack. The first non-vanishing term in an expansion
of the amplitude in powers of $\alpha'$ is then:
\begin{equation}
\mathcal{A}^{(2)}(\lambda_L \lambda_R \rightarrow BB) = \frac{N V_c}{N V_c + N' V'_c} \mathcal{A}^{66}
\end{equation}
where $\mathcal{A}^{66}$ is given by~(\ref{AALLexp}) (remember that in~(\ref{AALLexp}) the
Goldstino is just the U(1) gaugino). Using the relation~(\ref{gym_relation}) we obtain
finally:
\begin{eqnarray}
\mathcal{A}^{(2)}(\lambda_L \lambda_R \rightarrow BB) & = & - i (\frac{2 \pi^2
{\alpha'}^2 g_{YM}g'_{YM}V^{\frac{1}{2}}_c {V'}^{\frac{1}{2}}_{c}}{N
V_c + N' V'_c} ) \nonumber \\ & \times & (2\pi)^{4}
\delta^{(4)}(\sum_{i}k_{i}) \Tr({\lambda^{3} \lambda^{4}})
K_{GB}(1,2,3,4) \label{AALLexpII}
\end{eqnarray}
Here the Chan-Paton matrices $\lambda^3$ and $\lambda^4$ are in the
Lie-Algebra of $U(N)$ and the kinematic factor $K_{GB}(1,2,3,4)$ is given
by~(\ref{AA_kinematic_factor}). The result is unchanged if we
replace $B$ by $B'$, the gauge boson of the $D6'$ stack, except of course that
the Chan-Paton matrices then lie in the Lie-Algebra of $U(N')$.

In the same way we can obtain interaction amplitudes of the Goldstino
with adjoint scalars and fermions of $U(N)$. The Goldstino again
``appears'' as a gaugino with a non-canonical normalization and the
amplitudes $\mathcal{A}^{(2)}(\lambda_L \lambda_R \rightarrow \phi^{(i)} \bar{\phi}^{(j)})$,
$\mathcal{A}^{(2)}(\lambda_L \lambda_R \rightarrow f_L \bar{f}_R)$ and
$\mathcal{A}^{(2)}(\lambda_L \lambda_L \rightarrow f_L f_L)$ are
obtained by multiplying respectively~(\ref{GGSS_untwisted}),~(\ref{GGFF_untwisted})
and~(\ref{GGFF_untwisted_b})
with the factor $\frac{N V_c}{N V_c + N' V'_c}$:
\begin{eqnarray}
\mathcal{A}^{(2)}(\lambda_{L} \lambda_{R} \rightarrow \phi^{(i)}
\bar{\phi}^{(j)}) & = & - i  (\frac{2 \pi^2 {\alpha'}^2
g_{YM}g'_{YM}V^{\frac{1}{2}}_c {V'}^{\frac{1}{2}}_{c}}{N V_c + N'
V'_c} ) \nonumber \\ & \times &
(2\pi)^{4} \delta^{(4)}(\sum_{i}k_{i}) \Tr({\lambda^{3} \lambda^{4}}) \delta^{ij}
K_S(1,2,3,4) \label{GGSS_untwistedII} \\
\mathcal{A}^{(2)}(\lambda_{L} \lambda_{R} \rightarrow  f_{L} \bar{f}_{R}) & = &
- 2 i (\frac{2 \pi^2 {\alpha'}^2 g_{YM}g'_{YM}V^{\frac{1}{2}}_c
{V'}^{\frac{1}{2}}_{c}}{N V_c + N' V'_c} ) \nonumber \\ & \times &
(2\pi)^{4} \delta^{(4)}(\sum_{i}k_{i}) Tr({\lambda^{3} \lambda^{4}})
K_F(1,2,3,4) \label{GGFF_untwistedII} \\
\mathcal{A}^{(2)}(\lambda_L \lambda_L \rightarrow f_L f_L ) & = & -2i (\frac{2
  \pi^2 {\alpha'}^2 g_{YM} g'_{YM} {V_c}^{\frac{1}{2}} {V'_c}^{\frac{1}{2}}}{N
  V_c + N' V'_c}) \nonumber \\ & \times & (2\pi)^4 \delta^4(\sum_i k_i)
\Tr(\lambda^3 \lambda^4) K'_F(1,2,3,4) \label{GGFF_untwistedII_b}
\end{eqnarray}
The kinematic factors $K_S(1,2,3,4)$, $K_F(1,2,3,4)$ and $K'_F(1,2,3,4)$ are given
respectively by~(\ref{SS_kinematic_factor}),~(\ref{four_fermions1}) and~(\ref{four_fermions1_b}). Again these results are valid for adjoint scalars and fermions of both the D$6$ and D$6'$ stacks, provided that the
Chan-Paton factors lie, respectively, in the Lie-Algebra of $U(N)$ and
$U(N')$. Notice also that in the limit $\phi_k \rightarrow 0$ where the two stacks become
coincident, we recover the results of section~\ref{single_stack}.

\section{The low-energy effective action} \label{effective_action}
\setcounter{equation}{0}

We are now ready to compare the string computations with the effective
low-energy quantum field theory. We first concentrate on the twisted
sector that corresponds to strings localized on brane intersections and
then study the untwisted sector corresponding to strings ending on a
single D-brane stack.

\subsection{States on D-brane intersections} \label{effective_action_twisted}

As discussed before, on the intersection there is a massless left-handed
fermion transforming in the $(N,\bar{N'})$ representation of $U(N) \times
U(N')$ (the $66'$ string) and its CPT conjugate right-handed
(anti-)fermion transforming in $(\bar{N},N')$ (the $6'6$ string). The
vertex operators for these states are given in (\ref{FermionVertex}). Let
us define $f_{iJ}$ the quantum field which absorbs the left-handed
fermion and creates the right-handed anti-fermion. This is a left-handed
Weyl spinor transforming in $(N, \bar{N'})$: $i$ is an index in the
fundamental of $U(N)$ and $J$ an index in the anti-fundamental of $U(N')$.

There is also a potentially massless scalar transforming in
$(N,\bar{N}')$ and its CPT (complex) conjugate. Accordingly, let us
define $\phi_{iJ}$ the quantum field which annihilates this scalar and
creates its CPT conjugate. Of course, this scalar will appear in the
effective theory only if the choice of intersection angles makes it
massless, i.e. if the D-brane configuration is supersymmetric.

As mentioned before, we leave the coefficients of $S_6$ and $S_8$ (see appendix~\ref{appendixA}) undetermined. The latter does not contribute to four-point functions and therefore will not affect the following discussion. On the other hand, $S_6$ can not be discarded so easily. Indeed, the gauge bosons $A_\mu$ and $A'_{\mu}$ of $U(1)$ and $U(1)'$, respectively,  couple to fermions on the intersection via the usual Yang-Mills (YM) coupling:
\begin{equation}
\mathcal{L}_I = \frac{g_{YM}}{\sqrt{N}} f_{iJ} \sigma^\mu \bar{f}_{iJ} A_{\mu} - \frac{g'_{YM}}{\sqrt{N'}} f_{iJ} \sigma^\mu \bar{f}_{iJ} A'_{\mu}\ . \label{YM_coupling}
\end{equation}
Combining such a YM vertex with $S_6$ yields a contact term which has just the form of $S_3$. A similar problem occurs for the scalars on the intersection: $S_6$ together with the YM coupling results in a contact term undistinguishable from $S_5$. It might therefore seem that without determining $C_6$ we cannot obtain a definite prediction for the coefficients $C_3$ and $C_5$ in string theory. 
Fortunately, a more detailed analysis shows that these reducible contributions actually vanish. This is due to the fact that strings located on the intersection couple to the gauge bosons of $U(1)$ and $U(1)'$ with opposite charges, as can be seen from the relative minus sign in the two terms of~(\ref{YM_coupling}). As a result, the previously mentioned contact terms receive in fact two contributions which are equal and opposite and they cancel out. 

This cancellation is easiest to see in the ``string frame'' with Chan-Paton matrices normalized to $N$ ($N'$) for the $D6$ ($D6'$) stack. The tree-level low-energy effective Lagrangian for the $U(1)$ and $U(1)'$ vector multiplets is then:
\begin{equation}
\mathcal{L} = \frac{N V_c M_s^3}{g_s} \mathcal{L}^{D6} + \frac{N' V'_c M_s^3}{g_s} \mathcal{L}^{D6'} \label{string_frame}
\end{equation}
with
\begin{eqnarray}
\mathcal{L}^{D6} & = & - \frac{i}{2} g \sigma^\mu \overset{\leftrightarrow}{\partial_{\mu}} \bar{g} - \frac{1}{4} F^2 + \frac{C}{M_s^4}\,  \partial^{\alpha}g \sigma^{\mu} \partial^{\nu} \bar{g} \partial_{\alpha} F_{\mu \nu} + \cdots \\
\mathcal{L}^{D6'} & = & - \frac{i}{2} g' \sigma^\mu \overset{\leftrightarrow}{\partial_{\mu}} \bar{g'} - \frac{1}{4} {F'}^2 + \frac{C}{M_s^4} \, \partial^{\alpha}g' \sigma^{\mu} \partial^{\nu} \bar{g'} \partial_{\alpha} F'_{\mu \nu} + \cdots
\end{eqnarray}
Here $g$ and $g'$ are the gauginos of $U(1)$ and $U(1)'$, respectively,
and the dots indicate further terms that are not relevant for our
discussion. Notice that $\mathcal{L}^{D6}$ and $\mathcal{L}^{D6'}$ have
identical forms: all ``stack-dependant'' normalizations have been
absorbed in the fields and appear as overall factors of the two terms
in~(\ref{string_frame}); $g_s^{-1}$ is the contribution of the disc to
the genus expansion, while $V_c$ and $V'_c$ result from the
compactification of the internal dimensions of the two stacks. Also, all
string amplitudes are weighted by traces of products of Chan-Paton
matrices. For strings belonging to the $U(1)$ ($(U(1)'$) supermultiplet
and with our present normalization convention, these Chan-Paton factors
are just unit matrices and their trace contributes an overall factor $N$ ($N'$).

We can now re-express~(\ref{string_frame}) in terms of canonically
normalized fields by rescaling:
\begin{equation}
A_\mu \rightarrow (\frac{g_s}{NV_c M_s^3})^{\frac{1}{2}} A_\mu \qquad g \rightarrow (\frac{g_s}{NV_c M_s^3})^{\frac{1}{2}} g
\end{equation}  
and similarly for $A'_\mu$ and $g'$. Also, we have to use~(\ref{physical_goldstino}) to express the Goldstino as a linear combination of $g$ and $g'$. The Lagrangian then becomes
\begin{eqnarray}
\mathcal{L} & = & -\frac{i}{2} \lambda \sigma^{\mu} \overset{\leftrightarrow}{\partial_\mu} \bar{\lambda} - \frac{1}{4} F^2 - \frac{1}{4} {F'}^2 \nonumber \\
& + & \frac{1}{M_s^7} \frac{C g_s}{NV_c + N' V'_c} \frac{\sqrt{N}}{g_{YM}} (\partial^{\alpha} \lambda \sigma^{\mu} \partial^{\nu} \bar{\lambda}) \partial_{\alpha} F_{\mu \nu} \nonumber \\ & + & \frac{1}{M_s^7} \frac{C g_s}{N V_c + N' V'_c} \frac{\sqrt{N'}}{g'_{YM}} (\partial^{\alpha} \lambda' \sigma^{\mu} \partial^{\nu} \bar{\lambda}') \partial_{\alpha} F'_{\mu \nu} + \cdots
\end{eqnarray}
where we have used that $\frac{g_s}{M_s^3} = g^2_{YM} V_c = {g'}^2_{YM} V'_c$.
The last two terms will contribute to the action two copies of $S_6$, one for each abelian gauge
boson. Combining these with~(\ref{YM_coupling}) to build four-fermion
interactions of the same type as those generated by $S_3$, we obtain two
diagrams which contribute with opposite sign and cancel out. The same
reasoning applies if we replace the chiral fermions by their scalar
superpartners on the intersection (if they are present). We conclude that
$S_6$ will in fact never affect the determination of $C_3$ and $C_5$ and
thus for our purposes its coefficient $C_6$ is irrelevant.  

In the twisted case, we can exclude the term $S_1$ from the effective
action since there are is no adjoint matter on the intersection. Gauge
invariance also excludes the term $S_4$. Finally, as explained
in Section~\ref{section_twisted scalars}, conservation of internal helicity forbids
the interaction $S_7$. The full effective action has then the
form\footnote{Since $S_6$ and $S_8$ are irrelevant for the following discussion, we have not included them in~(\ref{twisted_effective_action}).}
\begin{equation}
S = S_0 + S_2 + S_3 + S_5\, , \label{twisted_effective_action}
\end{equation}
where $S_0$ is the model-independent coupling
\begin{equation}
S_0 = \int d^4 x (i \kappa^2 \lambda
\overset{\leftrightarrow}{\partial^{\mu}} \sigma^{\nu}
\bar{\lambda}) T_{\mu\nu}\, . \label{model_independent_coupling}
\end{equation}
In the present context, the relevant part of the energy-momentum tensor is
\begin{eqnarray}
T_{\mu\nu} & \equiv & T^{F}_{\mu\nu} + T^{S}_{\mu\nu} \nonumber \\ &
= & -\frac{i}{2} (f_{iJ} \sigma_{\mu} \overset{\leftrightarrow}{D_{\nu}} \bar{f}_{iJ}) + (D_{\mu}
\phi)_{iJ}^{\dagger} D_{\nu}\phi_{iJ} + (D_{\nu} \phi)_{iJ}^{\dagger}
D_{\mu}\phi_{iJ}
\end{eqnarray}
where it is understood that the scalar part $T^S_{\mu\nu}$ must be
included only if the scalar is massless.

The general form of the couplings $S_{i}$ is given in appendix
\ref{appendixA}. Using that the fields transform in bifundamentals,
we obtain:
\begin{eqnarray}
S_2 & = & C_{2} \kappa \int d^4 x (f_{iJ} \partial_\alpha
\lambda)D^{\alpha}\phi^{\dagger}_{iJ} + h.c. \label{S2} \label{coefficient_C2} \\
S_3 & = & C_3 2 \kappa^2 \int d^4 x (\bar{f_{iJ}} \partial^\mu
\bar{\lambda})(f_{iJ} \partial_\mu \lambda) \label{coefficient_C3} \\
S_5 & = & C_5 \kappa^2 \int d^{4}x\
(\phi_{iJ}^{\dagger} D_{\mu}\phi_{iJ} - (D_{\mu}\phi_{iJ})^{\dagger} \phi_{iJ})
i \partial_{\alpha} \lambda \sigma^{\mu} \partial^{\alpha}\bar{\lambda}
\end{eqnarray}
Here $C_3$ and $C_5$ are real while $C_2$ is in general complex. Again $S_2$ and $S_5$ are included only if there is a massless scalar in the spectrum. However, as we show below, the interaction $S_2$ may be absent even when a massless scalar is present. 

A necessary condition for $S_2$ to be present in the effective action is that the vertex operator
of the massless scalar appears in the fusion product of the vertex operators of the chiral fermion
and the Goldstino. We therefore start by considering OPEs of the vertex operators involved in~(\ref{S2}).
The OPE of the $66'$ string fermion vertex,
$\nu^{66'}_{-\frac{1}{2}}$, with that of the gaugino (\ref{GoldVertex})
gives:
\begin{equation}
\nu^{G}_{-\frac{1}{2}}(x_1) \nu^{66'}_{-\frac{1}{2}}(x_2)  =
x_{12}^{2\alpha' p_1 p_2 - \frac{\phi}{2\pi} + \frac{\epsilon-1}{2}}
\nu_{-1}^{S}(x_2) + \dots \label{fusion1a}
\end{equation}
where
\begin{equation}
\nu^{S}_{-1} =  e^{-\phi} \prod^{3}_{k=1} \sigma_{\epsilon_k}^k
e^{i(\epsilon_k(\frac{\phi_k}{\pi} - \frac{1}{2}) - \frac{1}{2})H_k}
e^{i(p_1 + p_2)X}\, .
\end{equation}
The string state created by the vertex $\nu^S_{-1}$ is a spacetime
scalar located on the intersection of the D-branes. Its mass can be
obtained directly by counting conformal weights:
\begin{equation}
\alpha' m^2 = \frac{\epsilon +1}{2} - \frac{\Phi}{2 \pi}\, ,
\label{scalar_mass}
\end{equation}
where $\Phi$ is given by~(\ref{sum_of_angles}).
This is the lightest state that can appear by ``fusing'' a left-handed
Goldstino with a massless fermion excitation of a $66'$ string. The dots
on the right-hand side of~(\ref{fusion1a}) denote scalars of higher mass.
In a simplified notation,~(\ref{fusion1a}) can be written as:
\begin{equation}
\nu^{G}_{-\frac{1}{2}} \times \nu^{66'}_{-\frac{1}{2}} = \nu^S_{-1} +
\dots \label{fusion1}
\end{equation}

In the same way, one can compute the fusion of a right-handed Goldstino
and a massless fermion excitation of $66'$ string. Now the emerging
states are spacetime vectors and we shall only need the lightest one:
\begin{equation}
\bar{\nu}^{G}_{-\frac{1}{2}} \times \nu^{66'}_{-\frac{1}{2}} =
\nu^V_{-1} + \dots \label{fusion2}
\end{equation}
where
\begin{equation}
\nu^V_{-1} =  e^{-\phi} (\zeta \psi) \prod^{3}_{k=1} \sigma_{\epsilon_k}^k
e^{i(\epsilon_k(\frac{\phi_k}{\pi} - \frac{1}{2}) + \frac{1}{2})H_k}
e^{i(p_1 + p_2)X}\, .
\label{vector_vertex}
\end{equation}
Here, $\zeta$ is a polarization vector which is given in terms of the
spinors appearing in $\bar{\nu}^G_{-\frac{1}{2}}$ and
$\nu^{66'}_{-\frac{1}{2}}$ by $\zeta^{\mu} = u^T_R C \gamma^{\mu} u_L$.
The string state corresponding to this vertex has a mass
\begin{equation}
\alpha' m^2 = \frac{1 - \epsilon}{2} + \frac{\Phi}{2 \pi}\, .
\label{vector_mass}
\end{equation}

Since the masses of the lightest resulting states~(\ref{scalar_mass})
and~(\ref{vector_mass}) depend on the signs of the rotations, we must
consider separately the cases $\epsilon=1$ (case I) and $\epsilon = -1$
(case II):

\begin{enumerate}
\item[I.]

The vertex operator $\nu^S_{-1}$ corresponds to the state
\begin{equation}
\bar{\psi}_{-\frac{1}{2} + \frac{\phi_1}{\pi}} \bar{\psi}_{-\frac{1}{2}
+ \frac{\phi_2}{\pi}} \bar{\psi}_{-\frac{1}{2} + \frac{\phi_3}{\pi}} |0;p>
\end{equation}
which is massive for all angles $0 \leq \phi_k \leq \frac{\pi}{2}$, as can be seen from~(\ref{scalar_mass}). We conclude that $S_2$ does not appear in the effective action and therefore $C_2 = 0$. To determine $C_3$, $C_5$ and $\kappa^2$
we should evaluate the amplitudes $\mathcal{A}(\lambda_L \lambda_R
\rightarrow f_L \bar{f}_R)$ and  $\mathcal{A}(\lambda_L \lambda_R
\rightarrow \phi \bar{\phi})$ in the effective quantum field theory and
compare them with the corresponding amplitudes in string theory. Starting
from the effective action~(\ref{twisted_effective_action}) with $C_2$ set
to zero, a short and straightforward computation gives the following QFT
amplitudes:
\begin{eqnarray}
\mathcal{A}^{QFT}(\lambda_L \lambda_R \rightarrow f_L \bar{f}_R) & =
& -2i \kappa^2 (2\pi)^4 \delta^{4}(\sum p_i) \delta_{IJ}\delta_{ij}
\frac{2tu + C_3 su}{2s}\hskip 1.2cm \\
\mathcal{A}^{QFT}(\lambda_L \lambda_R \rightarrow \phi \bar{\phi}) &
= & - i \kappa^2 (2\pi)^4 \delta^{4}(\sum p_i) \delta_{IJ}\delta_{ij}
\nonumber \\ & \times & (u_{1L}C \slash{p_4}u_{2R})(u - t + C_5 s)
\end{eqnarray}

These can be compared with the corresponding results of the string
calculations in the previous section,~(\ref{goldstino_totamplitude}),~(\ref{KFI})
and~(\ref{goldstino_scalar_amplitude}),\ (\ref{KSI}). We thus obtain:
\begin{equation}
\kappa^2 = \frac{2\pi^2 {\alpha'}^2 g_{YM} g'_{YM} V^{\frac{1}{2}}_c
{V'}^{\frac{1}{2}}_c}{N V_c + N' V'_c} \qquad C_3 = 1 \qquad C_5 = 0
\end{equation}

The supersymmetry breaking scale is proportional to the square of the
string length and depends on the intersection angles only through the
four-dimensional Yang-Mills couplings.  More precisely, it is given by
the effective tensions $T_{3}$ and $T'_3$ of the 3-brane stacks obtained
upon compactification of the internal directions along the D6-branes~\cite{Antoniadis:2001pt}
\footnote{The constant $v^{4}$ appearing in ref.~\cite{Antoniadis:2001pt}
is related to
$\kappa^{2}$ according to $v^{4} = \frac{1}{\kappa^{2}}$.}:
\begin{equation}
\frac{1}{ 2 \kappa^{2}} = N T_{3} + N' T'_3 \quad T_{3} = \frac{1}{ 4
\pi^{2} {\alpha'}^{2} g_{YM}^{2}} \quad  T'_{3} = \frac{1}{ 4 \pi^{2}
{\alpha'}^{2} {g'}_{YM}^{2}} \label{kappa2_general}
\end{equation}

As a result, in case I, the leading effective interactions of the
Goldstino with massless fields on brane intersections are given simply by
the coupling to the energy-momentum tensor and the four-fermion
interaction (\ref{coefficient_C3}) with coefficient $C_3=1$.
Note that the interaction $S_2$ is absent even in the presence of massless
scalars~(\ref{mass_shell_conditions}) at the intersection, when some
combination of angles $\phi_i+\phi_j-\phi_k=0$ for $i\ne j\ne k\ne i$.

\item[II.]

From the fusion rule (\ref{fusion1}) we now obtain:
\begin{equation}
\nu^G_{-\frac{1}{2}} \times \nu^{66'}_{-\frac{1}{2}} = \nu_{-1} + \dots \label{fusion_caseII}
\end{equation}
where $\nu_{-1}$ corresponds to one of the scalars in~(\ref{massless_scalars}). In fact, it is the scalar
which becomes massless precisely when $\Phi=0$. For example, if $\vec{\epsilon}=(-,-,+)$
then $\nu_{-1} \equiv \nu^{(3)}_{-1}$ (see~(\ref{scalar_intersection})) and the scalar is just $|j=3>$ in
(\ref{massless_scalars}). This is the reason for the apparent
discontinuity in the kinematic factors~(\ref{KFII}) and~(\ref{KSII}) when
$\Phi \rightarrow 0$. Indeed, as long as $\Phi \neq 0$, the scalar appearing on the right-hand side
of~(\ref{fusion_caseII}) is massive and just as in case I there is no operator $S_2$ in the
effective action. We then have again
\begin{equation}
\frac{1}{2\kappa^2} = N T_3 + N' T'_3 \qquad C_3 = 1 \qquad C_5 = 0 \label{coefficients_II}
\end{equation}

By continuity we expect that~(\ref{coefficients_II}) also holds when
$\Phi \rightarrow 0$. But in this limit the scalar becomes massless and $S_2$ will
contribute to the QFT amplitudes $\mathcal{A}(\lambda_L \lambda_R
\rightarrow f_L \bar{f}_R)$ and $\mathcal{A}(\lambda_L \lambda_R
\rightarrow \phi \bar{\phi})$. This extra contribution is responsible for
the discontinuities. The QFT amplitudes then become:
\begin{eqnarray}
\mathcal{A}^{QFT}(\lambda_L \lambda_R \rightarrow f_L \bar{f}_R) & = & -2i
\kappa^2 (2\pi)^4 \delta^{4}(\sum p_i) \delta_{IJ}\delta_{ij}
\nonumber \\ & & \frac{1}{2s} \left(tu(2- \frac{|C_2|^2}{4}) + su(C_3 -
\frac{|C_2|^2}{4})\right)\hskip 1cm \\
\mathcal{A}^{QFT}(\lambda_L \lambda_R \rightarrow \phi \bar{\phi}) & = & -
i \kappa^2 (2\pi)^4 \delta^{4}(\sum p_i) \delta_{IJ}\delta_{ij}
\nonumber \\ & & (u_{1L} C \slash{p_4} u_{2R}) (u - t +
C_5 s + \frac{|C_2|^2}{4} t)
\end{eqnarray}

Comparing these expressions with the results of the string calculations~(\ref{goldstino_totamplitude}),~(\ref{KFII}) and (\ref{goldstino_scalar_amplitude}),~(\ref{KSII}) in the case $\Phi=0$ we obtain
\begin{equation}
\frac{1}{2\kappa^2} = N T_3 + N' T'_3 \qquad |C_2|^2 = 4  \qquad C_3
= 1 \qquad C_5 = 0
\end{equation}

Thus, in case II, the effective action involving fields on brane
intersections is the same as in case I when $\Phi \neq 0$. However, when
$\Phi=0$, there is a massless scalar in the spectrum which couples
to the Goldstino via $S_0$ and the additional 3-point interaction
$S_2$. Note that in this case the intersection preserves locally
${\cal N}=1$ supersymmetry, which pairs the massless scalar with the
fermion in a chiral supermultiplet.

\end{enumerate}

\subsection{States on single D-brane stacks} \label{effective_action_untwisted}

We now consider untwisted fields corresponding to excitations of open
strings with both ends on the same stack of D-branes. The spectrum consists of a $U(N)$ gauge boson, four helicity $\frac{1}{2}$ adjoint fermions and their CPT conjugates and three complex adjoint scalars.

The fermions are labeled by their internal helicities. In our
conventions, the left-handed (right-handed) fermions have negative
(positive) internal chirality. The vertex operators for two CPT
conjugate fermions are
\begin{eqnarray}
\nu^{(j)}_{-\frac{1}{2}} & = & e^{-\frac{\phi}{2}} \lambda^a u_{L\alpha}
\Theta^{\alpha} e^{-\frac{i}{2}(\epsilon_1 H_1 + \epsilon_2 H_2 +
\epsilon_3 H_3)} e^{ikX} \nonumber \\
\bar{\nu}^{(j)}_{-\frac{1}{2}} & = & e^{-\frac{\phi}{2}} \lambda^a u_{R\alpha}
\Theta^{\alpha} e^{\frac{i}{2}(\epsilon_1 H_1 + \epsilon_2 H_2 +
\epsilon_3 H_3)} e^{ikX}
\end{eqnarray}
Here, the parameters $\epsilon_k = \pm 1$ must be chosen such that
$\epsilon_1 \epsilon_2 \epsilon_3 = 1$. The handedness of the spinors
is then imposed by the GSO projection. The index $j$ labels the
internal helicities in the following way:
\begin{equation}
j = \begin{cases} 0 & \text{if $\vec{\epsilon}=(+,+,+)$} \\ 1 &
\text{if $\vec{\epsilon}=(+,-,-)$} \\ 2 & \text{if
$\vec{\epsilon}=(-,+,-)$} \\ 3 & \text{if $\vec{\epsilon}=(-,-,+)$}
\end{cases} \label{index_j_defined}
\end{equation}
We define $f^a_j$ the (left-handed) quantum field which annihilates the left-handed
fermion corresponding to $\nu^{(j)}_{-\frac{1}{2}}$ and creates its
CPT conjugate anti-fermion $\bar{\nu}^{(j)}_{-\frac{1}{2}}$. The gauge
index $a$ is in the adjoint representation.

In the same way, we label the massless scalars. The vertex operators are
\begin{eqnarray}
\nu^{(j)}_{-1} & = & e^{-\phi} \lambda^a e^{-i H_j} e^{ikX} \nonumber \\
\bar{\nu}^{(j)}_{-1} & = & e^{-\phi} \lambda^a e^{i H_j} e^{ikX}
\end{eqnarray}
where $j=1,2,3$. We define $\phi^{a}_j$ the complex scalar field which annihilates
$\nu^{(j)}_{-1}$ and creates $\bar{\nu}^{(j)}_{-1}$. 

The leading effective action takes the form
\begin{equation}
S = S_0 + S_1 + S_2 + S_3 + S_4 + S_5 \label{untwisted_effective_action}
\end{equation}
and the energy-momentum tensor is given by
\begin{equation}
T_{\mu\nu} = T^{GB}_{\mu\nu} + T^F_{\mu\nu} + T^{S}_{\mu\nu}
\end{equation}
where
\begin{eqnarray}
T^{GB}_{\mu\nu} & = & - \Tr ( F_{\nu\sigma} F^{\sigma}_{\ \mu} + 
\frac{\eta_{\mu \nu}}{4} F_{\alpha\beta} F^{\alpha\beta})
\\
T^{F}_{\mu\nu} & = & - \frac{i}{2} \Tr (f_{j} \sigma_{\mu} \overset{\leftrightarrow}{D_\nu}
\bar{f_{j}} ) \\
T^{S}_{\mu\nu} & = & \Tr((D_{\mu} \phi_j)^{\dagger} D_{\nu}\phi_j
+ (D_{\nu} \phi_j)^{\dagger} D_{\mu}\phi_j)
\end{eqnarray}

Besides $S_0$ involving $T_{\mu\nu}$, the additional interactions are
obtained from Appendix \ref{appendixA} in the special case where the
fields transform in the adjoint of $U(N)$:
\begin{eqnarray}
S_1 & = & \kappa \sum^3_{j=0} C_1^j \int d^4 x\ iF^{a}_{\mu\nu}
f^{a}_j\sigma^\mu \partial^{\nu}\bar{\lambda} + h.c. \label{coefficient_C1b} \\
S_2 & = & \kappa \sum^3_{j=1} C^j_{2}\int d^4 x (f^a_j \label{coefficient_C2b}
\partial_\alpha \lambda)\partial^{\alpha}\phi^{a \dagger}_{j} + h.c.
\label{general_S2} \\
S_3 & = & \sum^3_{j=0} C^j_3 2 \kappa^2 \int d^4 x (\bar{f^a_{j}}
\partial^\mu \bar{\lambda})(f^a_{j} \partial_\mu \lambda) \\
S_4 & = & \sum_{j=0}^{3} C^j_4 \kappa^2 \int d^4 x (\bar{f}^a_j \bar{f}^a_j)(\partial_{\mu} \lambda \partial^{\mu} \lambda) + h.c.  \label{four_fermion_interactionII} \\
S_5 & = & \kappa^{2} \sum^3_{j=1} C^j_5 \int d^{4}x\
({\phi^{a}_{j}}^\dagger D_{\mu}\phi^{a}_{j} -
(D_{\mu}\phi^{a}_{j})^{\dagger} \phi^{a}_{j})
i \partial_{\alpha} \lambda \sigma^{\mu}
\partial^{\alpha}\bar{\lambda}
\end{eqnarray}
The interaction $S_4$ can only appear if the fermions have internal helicities opposite to the Goldstinos, as we have anticipated in~(\ref{four_fermion_interactionII}).

All untwisted vertex operators can be obtained from the corresponding
twisted expressions by taking the limit $\phi_k \rightarrow 0$. Starting
with the vertex operators $\nu^{66'}_{-\frac{1}{2}}$ and
$\bar{\nu}^{6'6}_{-\frac{1}{2}}$ and taking the limit $\phi_k \rightarrow
0$ (using that the bosonic twist $\sigma^k$ at vanishing angle is the
identity operator), we obtain, respectively, the fermion vertices $\nu^{(j)}_{-\frac{1}{2}}$ and $\bar{\nu}^{(j)}_{-\frac{1}{2}}$.

More precisely, if we start from case I we obtain $j=0$, i.e. fermions with
internal helicities $(+,+,+)$ and $(-,-,-)$. In this
limit, the fusion rule~(\ref{fusion1}) and the mass formula~(\ref{scalar_mass}) imply that the intermediate scalars are massive and therefore the interaction $S_2$ is absent. On the other hand,~(\ref{fusion2}) and~(\ref{vector_mass}) show that a left-handed fermion can combine with a right-handed Goldstino to yield a gauge boson. This implies that a new interaction $S_1$ may be present
and should be taken into account.

In case II, starting from the vertex $\nu^{66'}_{\frac{1}{2}}$ and taking the limit $\phi_k \rightarrow 0$ we obtain $\nu^{(j)}_{-\frac{1}{2}}$ with $j=1,2,3$. According to the fusion rules, these fermions can couple with Goldstinos to yield massless scalars but not
gauge bosons. Moreover, the fermion field $f^a_{j}$ can only couple to
the scalar ${\phi^a_{j}}^\dagger$, as we have anticipated in (\ref{general_S2}).

We conclude that
\begin{equation}
C^{j}_1 = 0 \quad \text{for j=1,2,3} \qquad \rm{and} \qquad C^{0}_2 = 0\ .
\end{equation}
The remaining coefficients can be obtained by the same amplitudes
studied in the previous subsection. Using the action~(\ref{untwisted_effective_action}),
a straightforward calculation yields:
\begin{eqnarray}
\mathcal{A}^{QFT}(\lambda_L \lambda_R \rightarrow f^{j}_L \bar{f}^{j}_R)
& = & -2i \kappa^2 (2\pi)^4 \delta^4(\sum k_i)  \delta^{ab} \nonumber \\ & \times & \frac{1}{2s} \biggl( tu (2-\frac{|C^{j}_1|^2}{2} - \frac{|C^j_2|^2}{4}) +
su (C^j_3 - \frac{|C^j_2|^2}{4}) \biggr) \nonumber \\ \label{amplitude1} \\
\mathcal{A}^{QFT}(\lambda_L \lambda_R \rightarrow \phi^j \bar{\phi}^j) & =
& - i \kappa^2 (2\pi)^4 \delta^{4}(\sum_i k_i)  \delta^{ab} \nonumber
\\  &\times& (u_{1L}C \slash{k_4}u_{2R})(u - t + C^j_5 s +
\frac{|C^j_2|^2}{4} t ) \nonumber \\ \label{amplitude2}
\end{eqnarray}
Since the above amplitudes involve Goldstinos of opposite helicity, the
interaction $S_4$ does not contribute. Note also that in~(\ref{amplitude1}) $j=0,...,3$ while in~(\ref{amplitude2}) $j=1,2,3$. Comparing these expressions with the string results~(\ref{GGSS_untwistedII})
and~(\ref{GGFF_untwistedII}) and using~(\ref{kappa2_general}) for the SUSY-breaking scale,
we obtain
\begin{equation}
|C^{0}_1|^2 = 2 \qquad |C^{1,2,3}_2|^2 = 4 \qquad C^{0,1,2,3}_3 = 1
\qquad C^{j}_5 = 0 \label{coefficients1}
\end{equation}

As an additional check we can use the effective action to compute the
QFT amplitude involving two Goldstinos and two gauge fields,
$\mathcal{A}(\lambda_L \lambda_R \rightarrow BB)$, and show that it
agrees with the string result~(\ref{AA_kinematic_factor}) and~(\ref{AALLexpII}).
Indeed, we obtain
\begin{equation}
\mathcal{A}^{QFT}(\lambda_L \lambda_R \rightarrow BB) = - i \kappa^2
(2\pi)^4 \delta^{4}(\sum_i k_i) \delta^{ab} K^{QFT}_{GB}
\end{equation}
with
\begin{equation}
K^{QFT}_{GB} = K_{GB}(1,2,3,4) + (1 - \sum^3_{j=0}
\frac{|C^j_{1}|^2}{2}) K'(1,2,3,4)\, ,
\end{equation}
where $K_{GB}(1,2,3,4)$ is the kinematic factor~(\ref{AA_kinematic_factor})
and $K'(1,2,3,4)$ is given by
\begin{equation}
\begin{split}
K'(1,2,3,4) = & u (\epsilon_3 \epsilon_4)
u^T_{1L}C\slash{k_4}u_{2R} - s (k_1 \epsilon_3)
u^T_{1L}C\slash{\epsilon_4}u_{2R} + \frac{s}{2}  u^T_{1L}C\slash{\epsilon_3}
\slash{\epsilon_4} \slash{k_4}u_{2R} \\ & + 2
u^T_{1L}C\slash{k_4}u_{2R}((k_3 \epsilon_4)(k_2 \epsilon_3) - (k_4
\epsilon_3)(k_2 \epsilon_4))\, .
\end{split}
\end{equation}
Using that $|C^0_1|^2 = 2$ and $C^{1,2,3}_1 = 0$ one finds perfect agreement with the
string computation.

To conclude this section, we still have to determine $C^j_4$. This can be
done by computing the amplitude~$\mathcal{A}(\lambda_L \lambda_L
\rightarrow f_L f_L )$, which is non-vanishing only if $j=0$ in
(\ref{index_j_defined}). We conclude:
\begin{equation}
C^{1,2,3}_4 = 0\, .
\end{equation}
A simple computation in QFT using the effective action
(\ref{untwisted_effective_action}) (actually only $S_1$ and $S_4$
contribute) then yields the amplitude:
\begin{equation}
\mathcal{A}^{QFT}(\lambda_L \lambda_L \rightarrow f^0_L f^0_L) = -2 i
\kappa^2 (2\pi)^4 \delta^4(\sum_i k_i) \delta^{ab} \frac{s}{2} 
(\frac{(\bar{C}_{1}^{0})^2}{2} + 2 C^0_4)\, ,
\end{equation}
where $\bar{C}^0_1$ is the complex conjugate of $C^0_1$. This result is
not enough to determine $C^0_4$ because $C^0_1$ has in principle a
phase ambiguity. However, if we assume that $C^0_1$ is real,
then we obtain simply that $C^0_4 = 0$ as can be seen by comparing
with~(\ref{GGFF_untwistedII_b}).

There are two reasons which motivate the assumption that $C^0_1$ (and
$C^j_2$) are real. First, it makes the interactions $S_1$ and $S_2$
invariant under spacetime parity transformations. Second, it allows to
write all order $\kappa$ interactions linear in the Goldstino as a
coupling to a supercurrent $J^{\mu}$ of the form
\begin{equation}
S = k \int d^4 x \, \kappa (\partial_{\mu} \lambda J^{\mu} + 
\partial_{\mu} \bar{\lambda} \bar{J}^{\mu})\, ,
\label{coupling_supercurrent}
\end{equation} 
where $k$ is a real constant.

In order to see this property, we consider first a supersymmetric D-brane
configuration. The $D6$ stack preserves the following supercharges:
\begin{equation}
Q = \mathcal{Q} + \beta_{\perp} \bar{\mathcal{Q}}\, ,
\end{equation}
where $\mathcal{Q}$ and $\mathcal{\bar{Q}}$ are the bulk supercharges of
the type II theory and $\beta_{\perp} = (\Gamma^5 \Gamma)(\Gamma^7 \Gamma)(\Gamma^9
\Gamma)$ implements a reflection in the directions transverse to the
D-branes which results from successive T-duality transformations. Here $\Gamma^M$ are the 
ten-dimensional gamma-matrices and $\Gamma \equiv \Gamma^{11}$ is the chirality operator. Similarly, the $D6'$ stack of D-branes preserves the supercharges
\begin{equation}
Q' = \mathcal{Q} + \beta'_{\perp}\bar{\mathcal{Q}}\, ,
\end{equation}
where $\beta'_{\perp}$ corresponds now to a reflection in the directions
transverse to the $D6'$ stack. Consider the supercharges
\begin{equation}
(Q_{\alpha, - - -},\ Q^{\dot{\alpha}}_{\ ,+ + +}) \label{supercharge_D6}
\end{equation}
and
\begin{equation}
({Q'}_{\alpha, - - -},\ {Q'}^{\dot{\alpha}}_{\ ,+ + +})\, ,
\label{supercharge_D6'} 
\end{equation}
where $\alpha, \dot{\alpha}$ are indices in the $(\frac{1}{2}, 0)$ and
$(0, \frac{1}{2})$ representations of the four-dimensional Lorentz group,
respectively. These supercharges form Majorana spinors in four
dimensions. The internal helicities are chosen to coincide with those of
the Goldstino and its CPT conjugate. 

Generically, the supercharges~(\ref{supercharge_D6})
and~(\ref{supercharge_D6'}) are distinct, in which case neither is
preserved by the D-brane intersection. However, if we choose the
intersection angles such that
\begin{equation}
(\beta_{\perp} \bar{\mathcal{Q}})_{\alpha, - - - } = (\beta'_{\perp} \bar{\mathcal{Q}})_{\alpha, - - -} \qquad (\beta_{\perp} \bar{\mathcal{Q}})^{\dot{\alpha}}_{\ ,+ + + } = (\beta'_{\perp} \bar{\mathcal{Q}})^{\dot{\alpha}}_{\ ,+ + +} 
\end{equation}
then the two supercharges coincide and the full spectrum is
supersymmetric. This happens precisely in case II (where not all angles
have the same sign) when the angles satisfy $\Phi=0$. The associated
conserved supercurrent is given by
\begin{equation}
J^{\mu} = J^{\mu}_{untwisted} + {J'}_{untwisted}^{\mu} + J_{twisted}^{\mu} + \dots \label{supercurrent}
\end{equation}
where
\begin{eqnarray}
J_{untwisted}^{\mu} & = &   \frac{i}{4} F^{a}_{\alpha \beta} [\sigma^{\alpha}, \bar{\sigma}^{\beta}] \sigma^\mu \bar{f}^a_{0} - \sqrt{2} \sum_{j=1}^{3} D_{\nu} \phi^{a \dagger}_{j } \sigma^{\nu} \bar{\sigma}^{\mu} f^a_j \nonumber \\
J^{\mu}_{twisted} & = & - \sqrt{2} D_{\nu} \phi^{\dagger}_{iJ} \sigma^{\nu} \bar{\sigma}^{\mu} f_{iJ} \label{supercurrent_explicit}
\end{eqnarray}
and ${J'}^{\mu}_{untwisted}$ has exactly the same form as
$J^{\mu}_{untwisted}$ with all fields replaced by primed fields
corresponding to states on the $D6'$-branes. The dots
in~(\ref{supercurrent}) denote terms that contain massive fields, which are
excluded from the effective action. Also, in~(\ref{supercurrent_explicit}), we have omitted  terms that become proportional to the equations of motion of the Goldstino when inserted in~(\ref{coupling_supercurrent}). Inserting~(\ref{supercurrent})
and~(\ref{supercurrent_explicit}) in~(\ref{coupling_supercurrent}) we
obtain
\begin{eqnarray}
S & = & k \int d^4 x\ \kappa \biggl( 2 \sqrt{2} \sum_{j=1}^3 (f^a_j 
\partial_{\mu} \lambda) D^{\mu} {\phi^a_j}^{\dagger} - 2i F^a_{\mu\nu}
\partial^{\nu} \lambda \sigma^{\mu} \bar{f}^a_0 \nonumber \\ & + & 2
\sqrt{2} (f_{iJ}\partial_{\mu} \lambda) D^{\mu} \phi^{\dagger}_{iJ} +
h.c. \biggr)\, .
\label{coupling_supercurrent_explicit}
\end{eqnarray}
This reproduces the order $\kappa$ couplings $S_1$ and $S_2$ obtained in
this section, provided that we set $k^2 = \frac{1}{2}$.

If we vary the angles, then typically the intersection breaks the
supersymmetries~(\ref{supercharge_D6}) and~(\ref{supercharge_D6'}). The
scalar superpartner of the chiral fermion becomes massive and the
coupling of the Goldstino to $J^{\mu}_{twisted}$ is removed from the
low-energy effective action. However, the spectrum on either D-brane
stack is still supersymmetric locally. On the $D6$-branes, the Goldstino
couples to the supercurrent $J^{\mu}_{untwisted}$ whose
supercharge~(\ref{supercharge_D6}) is preserved by the stack. Similarly,
the $D6'$-branes preserve the supercharge~(\ref{supercharge_D6'}) and the
Goldstino couples to the associated supercurrent
${J'}^{\mu}_{untwisted}$. 

To conclude, it is suggestive to postulate that the Goldstino couples
linearly to the supercurrent which has the same internal helicities as
itself. In our case, where the Goldstino and its CPT conjugate were
chosen with internal helicities $(- - -)$ and $(+ + +)$, this yields the
interaction~(\ref{coupling_supercurrent_explicit}). This
implies in particular that the coefficients $C_2$, $C_1^0$ and $C_2^j$ as
defined in~(\ref{coefficient_C2}),~(\ref{coefficient_C1b}) 
and~(\ref{coefficient_C2b}) are real. As a final remark, we note that this
coupling requires extended supersymmetry and is different from the
well-known coupling of the Goldstino to the ``spontaneously broken"
supercurrent (under which the Goldstino transforms non-linearly). Indeed,
this supercurrent would involve bosons and fermions of the same
supermultiplet and the Goldstino coupling would be proportional to the
corresponding mass-splitting. However, in our case, these mass-splittings
are strictly speaking infinite since there are no superpartners under the
non-linear supersymmmetry present on the branes.

Assuming the reality of all coefficients, the complete effective action is
given in Appendix \ref{appendixB}. It depends on the angles
only through the supersymmetry breaking scale. All coefficients are
independent of the angles. Of course, these results are only valid in the
limit $\alpha's, \alpha' t, \alpha' u <<
\Phi$. In this regime the energies are too small to ``resolve'' the
intersection and all interactions become insensitive to the precise
values of the angles. In order to probe the ``structure'' of the
intersection, we expect that it is necessary to consider energies (in string units) of the same order or higher than $\Phi$.
\section{Concluding remarks} \label{conclusion}

In this work, we classified all lower dimensional effective operators
describing interactions of the Goldstino with gauge fields, scalars and
chiral fermions, listed in Appendix A. Their strength is set by
appropriate powers of the supersymmetry breaking scale (or
equivalently, the Goldstino decay constant) times dimensionless
coefficients, which we have computed in string theory with intersecting
D-branes. The Goldstino decay constant is given by the total effective 3-brane
tension, while all couplings turn out to be universal constants,
independent from the values of the brane intersection angles. They are
summarized in Appendix B.

In the framework of low scale string theories with fundamental scale in
the TeV region and supersymmetric bulk, our analysis provides all
possible couplings of the Goldstino to Standard Model fields.
Two of them correspond to dimension six operators and can lead in
principle to the most dominant effects at low energy. They generate
three-point interactions of a single Goldstino with a chiral fermion
and a gauge field~(A.3) or a scalar~(A.4). In the absence of extra states,
such as adjoint fermions, the former involves in principle the
hypercharge and a fermion singlet such as the right-handed neutrino,
while (A.4) can exist for the Higgs and lepton doublets. Both couplings
seem to violate lepton number. A study of their effects is currently
under way~\cite{atz}.

\section*{Acknowledgements}

We would like to thank K. Benakli and F. Zwirner for useful discussions.
This work was supported in part by the European Commission under RTN
contract HPRN-CT-2000-00148, and in part by the INTAS contract
03-51-6346.

\setcounter{equation}{0}

\appendix
\setcounter{equation}{0}
\section{Goldstino couplings} \label{appendixA}

In this appendix, we summarize the full list of Goldstino couplings
consistent with non-linear supersymmetry up to order $\kappa^2$. The
model-independent coupling reads
\begin{equation}
S_0 =  \int d^4 x i (\kappa^2
\lambda\overset{\leftrightarrow}{\partial^{\mu}} \sigma^{\nu}
\bar{\lambda}) T_{\mu\nu}\, ,
\end{equation}
where $T_{\mu\nu}$ is the energy-momentum tensor\footnote{In the definition
of the energy-momentum tensor, the fields $\phi_i$ denote ``matter'' fields. Otherwise,
the symbol $\phi$ is used only for scalar fields.}
\begin{equation}
T^{\mu\nu} = \eta^{\mu\nu} \mathcal{L}_{SM} - \frac{\partial
\mathcal{L}_{SM}}{\partial(D_{\mu} \phi_{i})} D^{\nu}\phi_{i} + 2
\frac{\partial \mathcal{L}_{SM}}{\partial(F^{a}_{\mu\lambda})}
F^{a \nu}_{\lambda}\, .
\end{equation}
In addition, we have found eight model-dependent couplings:
\begin{eqnarray}
S_{1} & = & C_{1} \int d^4 x\ i \kappa F^{a}_{\mu\nu} f^{a}\sigma^\mu
\partial^{\nu}\bar{\lambda} + h.c. \label{O1} \\
S_{2} & = & C_{2} \int d^4 x\ \kappa M_{ij} ( f^{i}
\partial_{\alpha}\lambda) D^{\alpha} {\phi^j} + h.c. \label{O2} \\
S_{3} & = & C_{3} \int d^4 x\  \kappa^2 M_{ij} (\partial^{\mu}
\lambda f^{(1)}_i)(\partial_{\mu} \bar{\lambda} \bar{f}^{(2)}_j) +
h.c. \label{O3} \\
S_{4} & = & C_{4} \int d^4 x\ \kappa^2 M_{ij} (f^{(1)}_i
f^{(2)}_j)(\partial_{\mu}\lambda \partial^{\mu}\lambda) + h.c. \label{O4} \\
S_{5} & = & C_5 \int d^4 x\ \kappa^2 M_{ij}(D_{\mu} \phi^{(1)}_{i}
\phi^{(2)}_{j} -  \phi^{(1)}_{i} D_{\mu}\phi^{(2)}_{j} )
i \partial_{\alpha} \lambda
\sigma^{\mu} \partial^{\alpha} \bar{\lambda} + h.c. \label{O6} \\
S_{6} & = & C_{6} \int d^4 x\ \kappa^2 \partial^{\alpha} \lambda
\sigma^{\mu} \partial^{\nu} \bar{\lambda} \partial_{\alpha}
F_{\mu\nu} + h.c. \label{O7} \\
S_{7} & = & C_{7} \int d^4 x\ \kappa^{\frac{3}{2}} \phi^{(1)}_{i}
\phi^{(2)}_{j} M_{ij}\ \partial_{\mu} \lambda
J_{(\frac{1}{2},0)}^{\mu\nu} \partial_{\nu} \lambda + h.c. \label{O8}
\\
S_{8} & = & C_{8} \int d^4 x\ i \kappa^2 M_{ijk} \phi^{(1)}_{i}
\phi^{(2)}_{j} \phi^{(3)}_{k} (\partial_{\mu} \lambda
J_{(\frac{1}{2},0)}^{\mu\nu} \partial_{\nu} \lambda) + h.c. \label{O9}
\end{eqnarray}
In this list, we omitted operators linear in $\lambda$ and suppressed by
more than one power of $\kappa$. The coefficients $C_{i}$ are
dimensionless (in general complex) numbers, in principle of order one,  that depend on
the underlying fundamental theory. We have determined them for the case
of string theory on D-branes. The result is summarized in Appendix
\ref{appendixB}.

\section{Low-energy effective action} \label{appendixB}

In this appendix we give the low-energy effective action of string theory
describing Goldstino interactions on D6-branes. This action has the form
\begin{equation}
S_{eff} = S_0 + S_{twisted} + S_{untwisted} + S'_{untwisted}\, .
\label{total_effective_action}
\end{equation}
$S_0$ is the Goldstino interaction with the energy-momentum tensor:
\begin{equation}
S_0 = \int d^4 x  (i \kappa^2 \lambda
\overset{\leftrightarrow}{\partial^{\mu}} \sigma^{\nu}
\bar{\lambda}) T_{\mu\nu}\, ,
\end{equation}
where
\begin{equation}
T^{\mu\nu} = \eta^{\mu\nu} \mathcal{L}_{SM} - \frac{\partial
\mathcal{L}_{SM}}{\partial(D_{\mu} \phi_{i})} D^{\nu}\phi_{i} + 2
\frac{\partial \mathcal{L}_{SM}}{\partial(F^{a}_{\mu\lambda})}
F^{a \nu}_{\lambda}
\end{equation}
and $\kappa^2$ is given in terms of the effective $3$-brane tensions as
\begin{equation}
\frac{1}{2 \kappa^2} = N T_3 + N' T'_3 \ .
\end{equation}

$S_{twisted}$ contains the interactions of the Goldstino with the matter
localized on the intersection of two $D6$-brane stacks. It is given by
\begin{equation}
S_{twisted} = 2 \kappa^2 \int d^4 x (\bar{f}_{iJ} \partial_{\mu}
\bar{\lambda})(f_{iJ} \partial^{\mu} \lambda) + 2 \kappa ( \int d^4 x
(f_{iJ} \partial_{\alpha}\lambda) D^{\alpha}\phi^{\dagger}_{iJ} + h.c. )
\label{stwist}
\end{equation}
Here $f_{iJ}$ is the chiral fermion on the intersection and $\phi^\dagger_{iJ}$ is
its superpartner with respect to the supercharge~(\ref{supercharge_D6}), (\ref{supercharge_D6'}).
The precise definitions of the quantum fields are given in
Section \ref{effective_action_twisted}. It is
understood that the second term should be included only if the scalar involved
in the interaction is massless, i.e. whenever the sum of the relative
rotation angles $\Phi = 0$ in (\ref{sum_of_angles}) which is the condition for
the supersymmetry~(\ref{supercharge_D6}),~(\ref{supercharge_D6'}) to be preserved by
the intersection. Only the
four-fermion interaction is generically present.

The remaining two contributions to $S_{eff}$ in~(\ref{total_effective_action})
contain the interactions with matter on the
D$6$ and D$6'$ stacks, respectively:
\begin{eqnarray}
S_{untwisted} & = & \sqrt{2} i \kappa \int d^4 x F^a_{\mu\nu} f^a_{0}
\sigma^{\mu} \partial^{\nu}\bar{\lambda} \nonumber\\  & + & 2 \kappa
\sum^{3}_{j=1} \int d^4 x (f^a_{j} \partial_{\mu}
\lambda)D^{\mu} {\phi^a_{j}}^\dagger \label{suntwist} \\ 
& + & 2 \kappa^2 \sum^{3}_{j=0} \int
d^4 x (\bar{f}^a_{j} \partial_{\mu} \bar{\lambda})(f^a_{j}
\partial^{\mu} \lambda)\nonumber
\end{eqnarray}

The action $S'_{untwisted}$ has exactly the same form as
$S_{untwisted}$. For the precise definitions of the quantum fields see
Section \ref{effective_action_untwisted}.

In our analysis, we left undetermined the coefficient of the five-point
function $C_8$, as well as $C_6$ which may exist only for abelian gauge
fields. All other couplings that do not appear in~(\ref{stwist}) and~{suntwist}),
such as $C_4$, $C_5$ and $C_7$, are vanishing.
We have also argued that $C_1$ and $C_2$ are real. In this case, the
terms linear in the Goldstino correspond to its coupling to the
supercurrent with the same internal helicities as itself. This coupling
is given by
\begin{equation}
S_{linear} = \frac{\kappa}{\sqrt{2}} \int d^4 x (\partial_{\mu}\lambda J^{\mu} + \partial_{\mu} \bar{\lambda} \bar{J}^{\mu})
\end{equation}
with $J^{\mu}$ and $\bar{J}^{\mu}$ the left- and right-handed part of
the supercurrent, respectively. In our conventions, where the Goldstino
and its CPT conjugate were chosen to have internal helicities $(- - -)$
and $(+ + +)$,  the massless part of the supercurrent is given
in~(\ref{supercurrent}) and~(\ref{supercurrent_explicit}).

\end{document}